\documentstyle[12pt]{article}

\newcommand{\bd}{\begin{document}}
\newcommand{\ed}{\end{document}}
\newcommand{\bc}{\begin{center}}
\newcommand{\ec}{\end{center}}
\newcommand{\be}{\begin{eqnarray}}
\newcommand{\ee}{\end{eqnarray}}
\renewcommand{\thefootnote}{\alph{footnote}}
\newcommand{\se}{\section}
\newcommand{\sse}{\subsection}
\newcommand{\bi}{\bibitem}
\def\figcap{\section*{Figure Captions\markboth
     {FIGURECAPTIONS}{FIGURECAPTIONS}}\list
     {Figure \arabic{enumi}:\hfill}{\settowidth\labelwidth{Figure 999:}
     \leftmargin\labelwidth
     \advance\leftmargin\labelsep\usecounter{enumi}}}
\let\endfigcap\endlist \relax

\begin{document}

\begin{titlepage}

 \vskip 0.5in
 \null
\begin{center}
 \vspace{.15in}
{\LARGE {\bf
Exclusive Decays of $B\to K^{(*)}l^+l^-$ in the PQCD}
}\\
\vspace{1.0cm}  \par
 \vskip 2.1em
 {\large
  \begin{tabular}[t]{c}
{\bf Chuan-Hung Chen$^a$ and C.~Q.~Geng$^b$}
\\
\\
       {\sl ${}^a$Department of Physics, National Cheng Kung University}
\\   {\sl  $\ $Tainan, Taiwan,  Republic of China }
\\
\\
{\sl ${}^b$Department of Physics, National Tsing Hua University}
\\  {\sl  $\ $ Hsinchu, Taiwan, Republic of China }
\\
   \end{tabular}}
 \par \vskip 5.3em

\date{\today}
 {\Large\bf Abstract}
\end{center}

We study the exclusive decays of $B\to K^{(*)}l^+l^-$
within the framework of the PQCD.
We obtain the form factors for the  $B\to K^{(*)}$ transitions
in all allowed values of $q^2$, which agree with
 the lattice results.
We find that our
distributions of the decay
rates and leptonic asymmetries
are consistent with that given
in the other QCD models in the literature.

\end{titlepage}

\section{Introduction}

The recent CLEO measurement of the radiative $b\to s\gamma $ decay \cite
{cleo} has motivated theorists to study exclusive rare B meson decays such
as $B\to K^{(*)}l^{+}l^{-}$ \cite{Bmeson}. In the standard model, these rare
decays occur at loop level and provide us with information on the parameters
of the Cabibbo-Kobayashi-Maskawa (CKM) matrix elements \cite{ckm} as well as
various hadronic form factors. In this paper, we examine the decays of $B\to
(K,K^{*})l^{+}l^{-}$ within the framework of the perturbative QCD (PQCD).

The calculations of matrix elements for exclusive hadron decays
can be performed in the PQCD approach developed by Brodsky-Lepage
(BL) \cite{BL}. The application to B meson decays was first
carried out in Refs. \cite{SHB} and \cite{BD}. In the BL
formalism, the nonperturbative part is expressed as the hardon
wave functions which could be determined via various QCD models
such as the QCD sum rule method or lattice gauge theory and the
transition amplitude is factorized into the convolution of hadron
wave functions and the hard amplitude of the constituent quarks.
However, with the BL approach, the nonperturbative effects appear
 \cite{IS,RAD} if one of the constituent quarks carries nearly
all the momentum of hadron. To solve the problem, Li and Sterman
\cite{LS} proposed by including the transverse momentum of
constituent quark, $k_{T}$, and the Sudakov form factor to the
wave functions to suppress the soft contributions from higher
order corrections. In terms of the parameter $b$ with $b$ being
the conjugate variable of $k_{T}$, they also showed that the
effects can be also expressed as that in the BL factorization
formalism.

The modified PQCD factorization theorem for exclusive heavy meson decays has
been developed some time ago \cite{yl,ly,CL} and applied to nonleptonic $%
B\to D^{(*)}\pi (\rho )$ \cite{yl}, penguin induced radiation
$B\rightarrow K^{*}\gamma $ \cite{LL}, and $B\rightarrow KK$
\cite{ChenL} decays. These decays involve three scales: the
$M_{W}$ scale as the initial condition of renormalization-group
(RG) equation, the typical scale $t$ which reflects the specific
dynamics of the heavy meson decays, and the factorization scale
$1/b$.
 Above the factorization scale,
there are two large logarithms $\ln (M_{W}/t)$ and $\ln (tb)$,
generated from radiative corrections. The former gives the
evolution from $M_{W}$ down to $t$
described by the Wilson coefficient (WC), while the latter from $t$ to $1/b$%
. There also exist double logarithms $\ln ^{2}(Pb)$ arising from the
radiative correction to the meson wave function, where $P$ is the dominant
light-cone component of a meson momentum. Resuming these double logarithms
leads to a Sudakov form factor of $\exp (-s(P,b))$ which suppresses the
long-distance contributions in the large $b$ region such that the
applicability of the PQCD around the energy scale of the bottom quark mass
could be guaranteed.

The typical three-scale factorization formula is generally written as the
following convolution product
\begin{equation}
C\left( t\right) \otimes H\left( t\right) \otimes \phi \left( x,b\right)
\otimes \exp \left( -s\left( P,b\right) -2\int_{1/b}^{t}\frac{d\bar{\mu}}{%
\bar{\mu}}\gamma \left( \alpha _{s}\left( \bar{\mu}\right) \right) \right)
\label{FT}
\end{equation}
where $C(t)$, $H(t)$ and $\phi ( x,b)$ denote the WC, hard decay amplitude
and non-perturbative wave function, respectively, and the quark anomalous
dimension $\gamma =-\alpha _{s}/\pi $ is evaluated from $t$ to $1/b$. Except
$\phi( x,b) $ dictated by non-perturbative dynamics, all the convolution
factors in Eq. (\ref{FT}) are calculable. Note that differing from the
conventional factorization assumption (FA), the WC is also one of the
convolution parts in Eq. (\ref{FT}). Thus, the $\mu $ dependent problem
occurring in the FA could be solved naturally in the three-scale
factorization formula.

The paper is organized as follows. In Sec.~2, we study the form factors in
the framework of the PQCD for the decays of $B\to K^{(*)}$ transitions. In
Sec.~3, we derive the forms of the differential decay rates and lepton
asymmetries for $B\to K^{(*)}l^+l^-$ based on the PQCD. In Sec.~4, we give
the numerical analysis. We will also compare our results in the PQCD
approach with that in the other QCD models. In Sec.~5, we present our
conclusions.

\section{Form factors in the framework of the PQCD}

In the decay of $B\to H l^+l^-$, the momentum of $B$ ($H$) in the light-cone
coordinate is chosen as $P_1=(P^+_{1(2)},P^-_{1(2)},{\bf 0}_{T})$, where $%
P_1^{\pm}=M_B/\sqrt{2}$ and $P_{2}^{\pm }=( E_{H}\pm P_{H}) /\sqrt{2}$ with $%
E_{H}=\left( M_{B}^{2}+M_{H}^{2}-q^{2}\right) /2M_{B}$, $P_{H}=\sqrt{%
E_{H}^{2}-P_{H}^{2}}$, and $q^{2}$ is the squared momentum transfer. We
define the momentum of the light valence quark in the $B$ meson as $k_{1}$
and use $x_{1}=k_{1}^{+}/P_{1}^{+}$ with $k_1^+$ and $k_{1T}$ being the plus
and transverse components of $k_1$, respectively. The two light valence
quarks in the $H$ meson carry the longitudinal momenta $x_{2}P_{2}$ and $%
(1-x_{2})P_{2}$, and transverse momenta ${\bf k}_{2T}$ and $-{\bf k}_{2T}$,
respectively. 

The Sudakov resummations of large logarithmic corrections lead to the
exponential forms of $\exp (-S_{B})$ and $\exp (-S_{H})$ for $B$ and $H$
wave functions, respectively, where
\begin{eqnarray}
S_{B}(t) &=&s(x_{1}P_{1}^{+},b_{1})+2\int_{1/b_{1}}^{t}\frac{d{\bar{\mu}}}{%
\bar{\mu}}\gamma (\alpha _{s}({\bar{\mu}}))\;,  \nonumber \\
S_{H}(t)
&=&s(x_{2}P_{2}^{+},b_{2})+s((1-x_{2})P_{2}^{+},b_{2})+2\int_{1/b_{2}}^{t}%
\frac{d{\bar{\mu}}}{\bar{\mu}}\gamma (\alpha _{s}({\bar{\mu}}))\;.
\label{sbk}
\end{eqnarray}
In Eq. (\ref{sbk}), $b_{1}$and $b_{2}$ represent the transverse momentum
extents of $B$ and $H$ and are conjugate to the parton transverse momenta $%
k_{1T}$ and $k_{2T}$, respectively. The form for $s$ is written as \cite
{CS,BS,LS}
\begin{equation}
s(Q,b)=\int_{1/b}^{Q}\frac{d\mu }{\mu }\left[ \ln \left( \frac{Q}{\mu }%
\right) A(\alpha _{s}(\mu ))+D(\alpha _{s}(\mu ))\right] \;,  \label{fsl}
\end{equation}
where the anomalous dimensions $A$ and $D$ calculated to the two and
one-loop levels, respectively, are given by
\begin{eqnarray}
A &=&C_{F}\frac{\alpha _{s}}{\pi }+\left[ \frac{67}{9}-\frac{\pi ^{2}}{3}-%
\frac{10}{27}f+\frac{2}{3}\beta _{0}\ln \left( \frac{e^{\gamma _{E}}}{2}%
\right) \right] \left( \frac{\alpha _{s}}{\pi }\right) ^{2}\;,  \nonumber \\
D &=&\frac{2}{3}\frac{\alpha _{s}}{\pi }\ln \left( \frac{e^{2\gamma _{E}-1}}{%
2}\right) \;,
\end{eqnarray}
with $C_{F}=4/3$ being a color factor, $f=4$ the active flavor number, and $%
\gamma_{E}$ the Euler constant. For the running coupling constant, we use
\begin{equation}
\alpha _{s}(\mu )=\frac{4\pi }{\beta _{0}\ln (\mu ^{2}/\Lambda _{{\rm QCD}%
}^{2})}
\end{equation}
with $\beta_{0}=(33-2f)/3$.

To get the transition elements of $B\rightarrow H\left( H=K,\ K^{*}\right) $
with various types of vertices, we parametrize them in terms of the relevant
form factors as follows:

\begin{eqnarray}
\left\langle K\left( P_{2}\right) |\ V_{\mu }\ |B\left( P_{1}\right)
\right\rangle &=&F_{1}\left( q^{2}\right) P_{\mu }+F_{2}\left( q^{2}\right)
q_{\mu },  \nonumber \\
\left\langle K\left( P_{2}\right) |\ T_{\mu \nu }q^{\nu }\ |B\left(
P_{1}\right) \right\rangle &=&F_{T}\left( q^{2}\right) \left( q^{2}P_{\mu
}-q\cdot Pq_{\mu }\right) ,  \nonumber \\
\left\langle K^{*}\left( P_{2},\varepsilon \right) |\ V_{\mu }\ |B\left(
P_{1}\right) \right\rangle &=&iV\left( q^{2}\right) \epsilon _{\mu \nu
\alpha \beta }\varepsilon ^{*\nu }P^{\alpha }q^{\beta },  \nonumber \\
\left\langle K^{*}\left( P_{2},\varepsilon \right) |\ A_{\mu }\ |B\left(
P_{1}\right) \right\rangle &=&A_{0}\left( q^{2}\right) \varepsilon _{\mu
}^{*}+\varepsilon ^{*}\cdot q\left[ A_{1}\left( q^{2}\right) P_{\mu
}+A_{2}\left( q^{2}\right) q_{\mu }\right] ,  \nonumber \\
\left\langle K^{*}\left( P_{2},\varepsilon \right) |\ T_{\mu \nu }q^{\nu }\
|B\left( P_{1}\right) \right\rangle &=&iT\left( q^{2}\right) \epsilon _{\mu
\nu \alpha \beta }\varepsilon ^{*\nu }P^{\alpha }q^{\beta },  \nonumber \\
\left\langle K^{*}\left( P_{2},\varepsilon \right) |\ T_{\mu \nu }^{5}q^{\nu
}\ |B\left( P_{1}\right) \right\rangle &=&-T_{0}\left( q^{2}\right)
\varepsilon _{\mu }^{*}-\varepsilon ^{*}\cdot q\left[ T_{1}\left(
q^{2}\right) P_{\mu }+T_{2}\left( q^{2}\right) q_{\mu }\right] ,  \label{ff}
\end{eqnarray}
with
\begin{equation}
T_{0}\left( q^{2}\right) +\left[ T_{1}\left( q^{2}\right) P\cdot
q+T_{2}\left( q^{2}\right) q^{2}\right] =0  \label{teq}
\end{equation}
where $V_{\mu }=\ \bar{s}\ \gamma _{\mu }\ b$, $A_{\mu }=\bar{s}\ \gamma
_{\mu }\gamma _{5}\ b$, $T_{\mu \nu }=\bar{s}\ i\sigma _{\mu \nu }b$, and $%
T_{\mu \nu }^{5}=\bar{s}\ i\sigma _{\mu \nu }\gamma _{5}\ b$. The
correspondences of our notation to that usually used in the
literature are shown in the Appendix. Using the PQCD factorization
formula, the components of form factors defined in Eq. (\ref{ff})
are found to be

\begin{eqnarray}
F_{1}\left( q^{2}\right) &=&-8\pi
C_{F}M_{B}^{2}\int_{0}^{1}[dx]\int_{0}^{\infty }b_{1}db_{1}b_{2}db_{2}\ \phi
_{B}\left( x_{1},b_{1}\right) \   \nonumber \\
&&\left\{ \left[ r_{K}^{\prime }\left( 2\alpha _{2K}+2\beta _{2K}-1\right)
\phi _{K}^{\prime }\left( x_{2},b_{2}\right) \right. \right.  \nonumber \\
&&\left. -\left( 1+\beta _{2K}+\left( 1+2r_{K}-s\right) \alpha _{2K}\right)
\phi _{K}\left( x_{2},b_{2}\right) \right] E_{K}\left( t_{e}^{\left(
1\right) }\right) h^{K}\left( x_{1},x_{2},b_{1},b_{2}\right)  \nonumber \\
&&+\left[ -2r_{K}^{\prime }\left( 1-\alpha _{1K}-\beta _{1K}\right) \phi
_{K}^{\prime }\left( x_{2},b_{2}\right) \right.  \nonumber \\
&&\left. \left. +\left( r_{K}\left( 1-\alpha _{1K}+\beta _{1K}\right) -s\
\beta _{1K}\right) \phi _{K}\left( x_{2},b_{2}\right) \right] E_{K}\left(
t_{e}^{\left( 2\right) }\right) h^{K}\left( x_{2},x_{1},b_{2},b_{1}\right)
\right\}  \label{k-1} \\
F_{2}\left( q^{2}\right) &=&-8\pi
C_{F}M_{B}^{2}\int_{0}^{1}[dx]\int_{0}^{\infty }b_{1}db_{1}b_{2}db_{2}\phi
_{B}\left( x_{1},b_{1}\right) \   \nonumber \\
&&\{[-r_{K}^{\prime }(1+2\alpha _{2K}-2\beta _{2K})\phi _{K}^{\prime
}(x_{2},b_{2})+(1+\beta _{2K}  \nonumber \\
&&+(1+2r_{K}-s)\alpha _{2K})\phi _{K}(x_{2},b_{2})]E_{K}\left( t_{e}^{\left(
1\right) }\right) h^{K}\left( x_{1},x_{2},b_{1},b_{2}\right)  \nonumber \\
&&+[2r_{K}^{\prime }\left( 1-\alpha _{1K}+\beta _{1K}\right) \phi
_{K}^{\prime }\left( x_{2},b_{2}\right) +(r_{K}(1-\alpha _{1K}-\beta _{1K})
\nonumber \\
&&\left. \left. -(2-s)\beta _{1K})\phi _{K}\left( x_{2},b_{2}\right) \right]
E_{K}\left( t_{e}^{\left( 2\right) }\right) h^{K}\left(
x_{2},x_{1},b_{2},b_{1}\right) \right\}  \label{k-2} \\
F_{T}\left( q^{2}\right) &=&-8\pi C_{F}M_{B}\int_{0}^{1}[dx]\int_{0}^{\infty
}b_{1}db_{1}b_{2}db_{2}\phi _{B}\left( x_{1},b_{1}\right)  \nonumber \\
&&\times \left\{ \left[ -\alpha _{2K}r_{K}^{\prime }\phi _{K}^{\prime
}\left( x_{2},b_{2}\right) +\left( 1-2\beta _{2K}\right) \phi _{K}\left(
x_{2},b_{2}\right) \right] E_{K}\left( t_{e}^{\left( 1\right) }\right)
h^{K}\left( x_{1},x_{2},b_{1},b_{2}\right) \right.  \nonumber \\
&&\left. +\left[ 2r_{K}^{\prime }\left( 1-\alpha _{1K}\right) \phi
_{K}^{\prime }\left( x_{2},b_{2}\right) -\beta _{1K}\phi _{K}\left(
x_{2},b_{2}\right) \right] E_{K}\left( t_{e}^{\left( 2\right) }\right)
h^{K}\left( x_{2},x_{1},b_{2},b_{1}\right) \right\} \,,  \label{k-3}
\end{eqnarray}
\begin{eqnarray}
V\left( q^{2}\right) &=&8\pi C_{F}M_{B}\int_{0}^{1}[dx]\int_{0}^{\infty
}b_{1}db_{1}b_{2}db_{2}\ \phi _{B}\left( x_{1},b_{1}\right) \phi
_{K^{*}}\left( x_{2},b_{2}\right)  \nonumber \\
&&\times \left\{ \left[ 1-2\beta _{2K^{*}}-\sqrt{r_{K^{*}}}\alpha
_{2K^{*}}\right] E_{K^{*}}\left( t_{e}^{\left( 1\right) }\right)
h^{K^{*}}\left( x_{1},x_{2},b_{1},b_{2}\right) \right.  \nonumber \\
&&\left. +\left[ \sqrt{r_{K^{*}}}\left( 1-\alpha _{1K^{*}}\right) \right]
E_{K^{*}}\left( t_{e}^{\left( 2\right) }\right) h^{K^{*}}\left(
x_{2},x_{1},b_{2},b_{1}\right) \right\} ,  \label{v} \\
A_{0}\left( q^{2}\right) &=&8\pi
C_{F}M_{B}^{3}\int_{0}^{1}[dx]\int_{0}^{\infty }b_{1}db_{1}b_{2}db_{2}\phi
_{B}\left( x_{1},b_{1}\right) \phi _{K^{*}}\left( x_{2},b_{2}\right)
\nonumber \\
&&\times \left\{ \left[ \left( 1+r_{K^{*}}-s\right) \left( 1-2\beta
_{2K^{*}}+\sqrt{r_{K^{*}}}\alpha _{2K^{*}}\right) \right. \right.  \nonumber
\\
&&\left. -4r_{K^{*}}\alpha _{2K^{*}}+2\sqrt{r_{K^{*}}}\left( 1+\beta
_{2K^{*}}\right) \right] E_{K^{*}}\left( t_{e}^{\left( 1\right) }\right)
h^{K^{*}}\left( x_{1},x_{2},b_{1},b_{2}\right)  \nonumber \\
&&\left. +\sqrt{r_{K^{*}}}\left[ \left( 1+r_{K^{*}}-s\right) \left( 1-\alpha
_{1K^{*}}\right) -2\beta _{1K^{*}}\right] E_{K^{*}}\left( t_{e}^{\left(
2\right) }\right) h^{K^{*}}\left( x_{2},x_{1},b_{2},b_{1}\right) \right\} ,
\label{a0} \\
A_{1}\left( q^{2}\right) &=&8\pi C_{F}M_{B}\int_{0}^{1}[dx]\int_{0}^{\infty
}b_{1}db_{1}b_{2}db_{2}\ \phi _{B}\left( x_{1},b_{1}\right) \phi
_{K^{*}}\left( x_{2},b_{2}\right)  \nonumber \\
&&\times \left\{ \left[ -1+2\beta _{2K^{*}}+\sqrt{r_{K^{*}}}\alpha
_{2K^{*}}\right] E_{K^{*}}\left( t_{e}^{\left( 1\right) }\right)
h^{K^{*}}\left( x_{1},x_{2},b_{2},b_{1}\right) \right.  \nonumber \\
&&+\left. \left[ \sqrt{r_{K^{*}}}\left( -1+\alpha _{1K^{*}}+2\beta
_{1K^{*}}\right) \right] E_{K^{*}}\left( t_{e}^{\left( 2\right) }\right)
h^{K^{*}}\left( x_{2},x_{1},b_{1},b_{2}\right) \right\} ,  \label{a1} \\
A_{2}\left( q^{2}\right) &=&8\pi C_{F}M_{B}\int_{0}^{1}[dx]\int_{0}^{\infty
}b_{1}db_{1}b_{2}db_{2}\ \phi _{B}\left( x_{1},b_{1}\right) \phi
_{K^{*}}\left( x_{2},b_{2}\right)  \nonumber \\
&&\times \left\{ \left[ 1-2\beta _{2K^{*}}-\sqrt{r_{K^{*}}}\alpha
_{2K^{*}}\right] E_{K^{*}}\left( t_{e}^{\left( 1\right) }\right)
h^{K^{*}}\left( x_{1},x_{2},b_{1},b_{2}\right) \right.  \nonumber \\
&&\left. +\left[ \sqrt{r_{K^{*}}}\left( 1-\alpha _{1K^{*}}+2\beta
_{1K^{*}}\right) \right] E_{K^{*}}\left( t_{e}^{\left( 2\right) }\right)
h^{K^{*}}\left( x_{2},x_{1},b_{2},b_{1}\right) \right\} ,  \label{a2} \\
T\left( q^{2}\right) &=&8\pi C_{F}M_{B}^{2}\int_{0}^{1}[dx]\int_{0}^{\infty
}b_{1}db_{1}b_{2}db_{2}\ \phi _{B}\left( x_{1},b_{1}\right) \phi
_{K^{*}}\left( x_{2},b_{2}\right)  \nonumber \\
&&\times \left\{ \left[ -\left( 1+\sqrt{r_{K^{*}}}\right) +2s\alpha
_{2K^{*}}+\left( 2\sqrt{r_{K^{*}}}-1\right) \left( \alpha _{2K^{*}}+\beta
_{2K^{*}}\right) \right] \right.  \nonumber \\
&&\times E_{K^{*}}\left( t_{e}^{\left( 1\right) }\right) h^{K^{*}}\left(
x_{1},x_{2},b_{1},b_{2}\right)  \nonumber \\
&&\left. +\sqrt{r_{K^{*}}}\left[ \alpha _{1K^{*}}+\beta _{1K^{*}}\left(
x_{1}\right) -1\right] E_{K^{*}}\left( t_{e}^{\left( 2\right) }\right)
h^{K^{*}}\left( x_{2},x_{1},b_{2},b_{1}\right) \right\} ,  \label{t} \\
T_{0}\left( q^{2}\right) &=&-8\pi
C_{F}M_{B}^{4}\int_{0}^{1}[dx]\int_{0}^{\infty }b_{1}db_{1}b_{2}db_{2}\ \phi
_{B}\left( x_{1},b_{1}\right) \phi _{K^{*}}\left( x_{2},b_{2}\right)
\nonumber \\
&&\times \left\{ \left[ \left( 1-r_{K^{*}}-s\right) \left( 1+\beta
_{2K^{*}}\right) +\sqrt{r_{K^{*}}}\left( 1-r_{K^{*}}+s\right) \right. \right.
\nonumber \\
&&-2\sqrt{r_{K^{*}}}\left( \left( 1-r_{K^{*}}\right) \left( \alpha
_{2K^{*}}+\beta _{2K^{*}}\right) +s\left( \beta _{2K^{*}}-\alpha
_{2K^{*}}\right) \right)  \nonumber \\
&&\left. -2r_{K^{*}}\alpha _{2K^{*}}+\left( 1+r_{K^{*}}-s\right) \left(
1-s\right) \alpha _{2K^{*}}\right] E_{K^{*}}\left( t_{e}^{\left( 1\right)
}\right) h^{K^{*}}\left( x_{1},x_{2},b_{1},b_{2}\right)  \nonumber \\
&&+\sqrt{r_{K^{*}}}\left[ \left( 1-\alpha _{1K^{*}}-\beta _{1K^{*}}\right)
\left( 1-r_{K^{*}}\right) \right.  \nonumber \\
&&\left. \left. -s\left( 1+\beta _{1K^{*}}-\alpha _{1K^{*}}\right) \right]
E_{K^{*}}\left( t_{e}^{\left( 2\right) }\right) h^{K^{*}}\left(
x_{2},x_{1},b_{2},b_{1}\right) \right\} ,  \label{t0} \\
T_{1}\left( q^{2}\right) &=&-8\pi
C_{F}M_{B}^{2}\int_{0}^{1}dx_{1}dx_{2}\int_{0}^{\infty
}b_{1}db_{1}b_{2}db_{2}\ \phi _{B}\left( x_{1},b_{1}\right) \phi
_{K^{*}}\left( x_{2},b_{2}\right)  \nonumber \\
&&\times \left\{ \left[ s\alpha _{2K^{*}}-\left( 1+\sqrt{r_{K^{*}}}\right)
-\left( 1-2\sqrt{r_{K^{*}}}\right) \left( \alpha _{2K^{*}}+\beta
_{2K^{*}}\right) \right] \right.  \nonumber \\
&&\times E_{K^{*}}\left( t_{e}^{\left( 1\right) }\right) h^{K^{*}}\left(
x_{1},x_{2},b_{1},b_{2}\right)  \nonumber \\
&&\left. -\sqrt{r_{K^{*}}}\left[ 1-\alpha _{1K^{*}}-\beta _{1K^{*}}\right]
E_{K^{*}}\left( t_{e}^{\left( 2\right) }\right) h^{K^{*}}\left(
x_{2},x_{1},b_{2},b_{1}\right) \right\} ,  \label{t1} \\
T_{2}\left( q^{2}\right) &=&-8\pi
C_{F}M_{B}^{2}\int_{0}^{1}dx_{1}dx_{2}\int_{0}^{\infty
}b_{1}db_{1}b_{2}db_{2}\ \phi _{B}\left( x_{1},b_{1}\right) \phi
_{K^{*}}\left( x_{2},b_{2}\right)  \nonumber \\
&&\times \left\{ \left[ \left( 2r_{K^{*}}-s\right) \alpha _{2K^{*}}+\left(
1+\alpha _{2K^{*}}+\beta _{2K^{*}}\right) \right. \right.  \nonumber \\
&&\left. -\sqrt{r_{K^{*}}}\left( 1+2\alpha _{2K^{*}}-2\beta _{2K^{*}}\right)
\right] E_{K^{*}}\left( t_{e}^{\left( 1\right) }\right) h^{K^{*}}\left(
x_{1},x_{2},b_{1},b_{2}\right)  \nonumber \\
&&\left. +\sqrt{r_{K^{*}}}\left[ 1-\alpha _{1K^{*}}+\beta _{1K^{*}}\right]
E_{K^{*}}\left( t_{e}^{\left( 2\right) }\right) h^{K^{*}}\left(
x_{2},x_{1},b_{2},b_{1}\right) \right\} ,  \label{t2}
\end{eqnarray}
where $\phi _{B}$, $\phi _{K}$ ($\phi _{K}^{\prime }$), and $\phi _{K^{*}}$
are the wave functions of $B$, pseudovector (pseudoscalar) of $K$, and $%
K^{*} $ mesons, respectively, the evolution factor are given by
\begin{equation}
E_{H}\left( t\right) =\alpha _{s}\left( t\right) \exp \left( -S_{B}\left(
t\right) -S_{H}\left( t\right) \right) ,  \label{ef}
\end{equation}
and the related kinematic variables are parametrized as
\begin{eqnarray}
\alpha _{1H} &=&-\frac{1}{\sqrt{\varphi _{H}}}x_{1},\quad \beta _{1H}=\frac{1%
}{2}\left( 1+\frac{1+r_{H}-s}{\sqrt{\varphi _{H}}}\right) x_{1},  \nonumber
\\
\beta _{2H} &=&-\frac{r_{H}}{\sqrt{\varphi _{H}}}x_{2},\quad \alpha _{2H}=%
\frac{1}{2}\left( 1+\frac{1+r_{H}-s}{\sqrt{\varphi _{H}}}\right) x_{2},
\nonumber \\
r_{H} &=&\frac{M_{H}^{2}}{M_{B}^{2}}\,,\ r_{H}^{\prime }\;=\;\frac{m_{0K}}{%
M_{B}}\,,\ s\;=\;\frac{q^{2}}{M_{B}^{2}}\,,\
\end{eqnarray}
with
\begin{eqnarray}
\varphi _{H} &=&\left( 1-r_{H}\right) ^{2}-2s\left( 1+r_{H}\right) +s^{2}\,,
\nonumber \\
m_{0K} &=&{\frac{M_{K}^{2}}{m_{s}+m_{d}}}\,.
\end{eqnarray}
The hard functions, $h^{H}$, are written as
\begin{eqnarray}
h^{H}\left( x_{1},x_{2},b_{1},b_{2}\right) &=&K_{0}\left( D_{H}\sqrt{%
x_{1}x_{2}}b_{1}\right) \left[ \theta \left( b_{1}-b_{2}\right) K_{0}\left(
D_{H}\sqrt{x_{2}}b_{1}\right) I_{0}\left( D_{H}\sqrt{x_{2}}b_{2}\right)
\right.  \nonumber \\
&&\left. +\theta \left( b_{2}-b_{1}\right) K_{0}\left( D_{H}\sqrt{x_{2}}%
b_{2}\right) I_{0}\left( D_{H}\sqrt{x_{2}}b_{1}\right) \right] ,
\end{eqnarray}
with
\[
D_{H}^{2}=\frac{M_{B}^{2}}{2}\left( 1+r_{H}-\frac{q^{2}}{M_{B}^{2}}+\sqrt{%
\varphi _{H}}\right) .
\]
The derivation of $h$, from the Fourier transformation of the lowest-order
hard decay amplitude, is similar to that for $B\rightarrow KK$ decays \cite
{ChenL}. The hard scales $t^{(1,2)}$ are chosen by
\begin{eqnarray}
t^{\left( 1\right) } &=&\max \left( \sqrt{x_{2}}D_{H},1/b_{1},1/b_{2}\right)
\,,  \nonumber \\
t^{\left( 2\right) } &=&\max \left( \sqrt{x_{1}}D_{H},1/b_{1},1/b_{2}\right)
\,.  \label{tscale}
\end{eqnarray}
The wave functions $\phi _{H}$ and $\phi _{K}^{\prime }$ are defined by \cite
{ChenL,KLS}
\begin{eqnarray}
\phi _{H}(x) &=&\int \frac{dy^{+}}{2\pi }e^{-ixP_{3}^{-}y^{+}}\frac{1}{2}%
\langle 0|{\bar{u}}(y^{+})\gamma ^{-}\gamma _{5}s(0)|H\rangle \;,  \nonumber
\\
\frac{m_{0K}}{P_{3}^{-}}\phi _{K}^{\prime }(x) &=&\int \frac{dy^{+}}{2\pi }%
e^{-ixP_{3}^{-}y^{+}}\frac{1}{2}\langle 0|{\bar{u}}(y^{+})\gamma
_{5}s(0)|K\rangle \;,
\end{eqnarray}
with the normalization conditions of
\[
\int_{0}^{1}dx\phi _{B}(x)=\int_{0}^{1}dx\phi _{H}(x)=\int_{0}^{1}dx\phi
_{K}^{\prime }(x)=\frac{f_{B\left( H\right) }}{2\sqrt{2N_{c}}}\;.
\]
We note that unlike the kaon case, we do not distinguish the pseudovector
and pseudoscalar components of the $B$ wave functions since the factor $%
M_{B}/(m_{b}+m_{d})$ is close to one. We also note that from Eqs. (\ref{t}) and (%
\ref{t1}) we obtain $T(0)=-T_{1}(0)$.

\section{Differential decay rates and lepton asymmetries}

The effective Hamiltonian of $b\rightarrow s\ l^{+}\ l^{-}$ is given by \cite
{Buras}

\begin{equation}
{\cal H}=\frac{G_{F}\alpha \lambda _{t}}{\sqrt{2}\pi }\left[ C_{8}\left( \mu
\right) \bar{s}_{L}\gamma _{\mu }b_{L}\ \bar{l}\gamma ^{\mu }l+C_{9}\bar{s}%
_{L}\gamma _{\mu }b_{L}\ \bar{l}\gamma ^{\mu }\gamma _{5}l-\frac{%
2m_{b}C_{7}\left( \mu \right) }{q^{2}}\bar{s}_{L}i\sigma _{\mu \nu }q^{\nu
}b_{R}\ \bar{l}\gamma ^{\mu }l\right]  \label{heff}
\end{equation}
where $C_{8}(\mu )$, $C_{9}$ and $C_{7}(\mu )$ are the WCs and their
expressions can be found in Ref. \cite{Buras} for the SM. Since the operator
associated with $C_{9}$ is not renormalized under the QCD, it is the only
one with the $\mu $ scale free. Besides the short-distance (SD)
contributions, the main effect on the branching ratio comes from c\={c}
resonant states such as $\Psi ,\Psi ^{\prime }\,...\,etc.,$ $i.e.$, the
long-distance (LD) contributions. In the literature \cite{DTP,LMS,AMM,OT,KS}%
, it has been suggested by combining FA and vector meson dominance (VMD)
approximation to estimate LD effects for the $B$ decays. Hence, including
the resonant effect (RE) and absorbing it to the related WC, we obtain the
effective WC of $C_{8}$ as
\begin{equation}
C_{8}^{eff}=C_{8}\left( \mu \right) +\left( 3C_{1}\left( \mu \right)
+C_{2}\left( \mu \right) \right) \left( h\left( x,s\right) +\frac{3}{\alpha }%
\sum_{j=\Psi ,\Psi ^{\prime }}k_{j}\frac{\pi \Gamma \left( j\rightarrow
l^{+}l^{-}\right) M_{j}}{q^{2}-M_{j}^{2}+iM_{j}\Gamma _{j}}\right) \,,
\label{effc8}
\end{equation}
where we have neglected the small Wilson coefficients, $h(x,s)$ describes
the one-loop matrix elements of operators $O_{1}=\bar{s}_{\alpha }\gamma
^{\mu }P_{L}b_{\beta }\ \bar{c}_{\beta }\gamma _{\mu }P_{L}c_{\alpha }$ and $%
O_{2}=\bar{s}\gamma ^{\mu }P_{L}b\ \bar{c}\gamma _{\mu }P_{L}c$ \cite{Buras}%
, $M_{j}$ ($\Gamma _{j}$) are the masses (widths) of intermediate states,
and the factors $k_{j}$ are phenomenological parameters for compensating the
approximations of FA and VMD and reproducing the correct branching ratios $%
Br\left( B\rightarrow J/\Psi X\rightarrow l^{+}l^{-}X\right) =Br\left(
B\rightarrow J/\Psi X\right) $ $\times $ $Br\left( J/\Psi \rightarrow
l^{+}l^{-}\right) $. In this paper, for simplicity, we take $k_{j}=-1/\left(
3C_{1}\left( \mu \right) +C_{2}\left( \mu \right) \right) $.

Using Eqs. (\ref{ff}) and (\ref{heff}), the transition amplitudes of $%
B\rightarrow (K,K^{*})\ l^{+}\ l^{-}$ are as follows
\begin{equation}
{\cal M}_{K}=\frac{G_{F}\alpha \lambda _{t}}{2\sqrt{2}\pi }\left\{ \left[
\left( F_{1}^{8}\left( q^{2}\right) -2m_{b}F_{T}^{7}\left( q^{2}\right)
\right) P_{\mu }\right] \ \bar{l}\gamma ^{\mu }l+\left[ F_{1}^{9}\left(
q^{2}\right) P_{\mu }+F_{2}^{9}\left( q^{2}\right) q_{\mu }\right] \ \bar{l}%
\gamma ^{\mu }\gamma _{5}l\right\}  \label{ampk}
\end{equation}
and
\begin{eqnarray}
{\cal M}_{K^{*}} &=&\frac{G_{F}\alpha \lambda _{t}}{2\sqrt{2}\pi }\left\{
\left[ i\left( V^{8}(q^{2})-\frac{2m_{b}}{q^{2}}T^{7*}(q^{2})\right)
\epsilon _{\mu \nu \alpha \beta }\varepsilon ^{*\nu }P^{\alpha }q^{\beta
}\right. \right.  \nonumber \\
&&\ \left. -\left( A_{0}^{8}(q^{2})-\frac{2m_{b}}{q^{2}}T_{0}^{7}(q^{2})%
\right) \varepsilon _{\mu }^{*}-\varepsilon ^{*}\cdot q\left(
A_{1}^{8}\left( q^{2}\right) -\frac{2m_{b}}{q^{2}}T_{1}^{7}\left(
q^{2}\right) \right) P_{\mu }\right] \ \bar{l}\gamma ^{\mu }l  \nonumber \\
&&+\left. \left[ iV^{9}(q^{2})\epsilon _{\mu \nu \alpha \beta }\varepsilon
^{*\nu }P^{\alpha }q^{\beta }-A_{0}^{9}(q^{2})\varepsilon _{\mu
}^{*}-\varepsilon ^{*}\cdot qA_{1}^{9}(q^{2})P_{\mu }\right] \ \bar{l}\gamma
^{\mu }\gamma _{5}l\right\} .  \label{ampk*}
\end{eqnarray}
In Eqs. (\ref{ampk}) and (\ref{ampk*}), we have included the WCs by
inserting them into Eq. (\ref{ef}) as
\begin{equation}
E_{H}^{j}\left( t\right) =C_{j}\left( \mu \right) \alpha _{s}\left( t\right)
\exp \left( -S_{B}\left( t\right) -S_{H}\left( t\right) \right) \,
\end{equation}
and the superscripts of form factors 
denote the associated WCs of $C_{8}^{eff}$, $C_{9}$, and $C_{7}$, with the
new definition of $E_{H}^{j}$ in Eqs. (\ref{k-1})-(\ref{t2}), respectively.

As usual, after integrating the angle dependent phase space, the
differential decay rates for $B\to H l^+l^-\ (H=K,K^*)$ are found to be

\begin{equation}
\frac{d\Gamma _{H}\left( q^{2}\right) }{ds}=\frac{G_{F}^{2}\alpha ^{2}\left|
\lambda _{t}\right| ^{2}M_{B}^{5}}{3\times 2^{9}\pi ^{5}}\sqrt{\varphi _{H}}%
\sqrt{1-\frac{4m_{l}^{2}}{q^{2}}}\left[ \left( 1+\frac{2m_{l}^{2}}{q^{2}}%
\right) \beta _{H}+12\frac{m_{l}^{2}}{M_{B}^{2}}\delta _{H}\right]
\label{dr}
\end{equation}
where 
\begin{eqnarray}
\beta _{K} &=&\varphi _{K}\left| F_{1}^{8}\left( q^{2}\right)
-2m_{b}F_{T}^{7}\left( q^{2}\right) \right| ^{2}+\varphi _{K}\left|
F_{1}^{9}\left( q^{2}\right) \right| ^{2},  \label{betak} \\
\delta _{K} &=&\left( 1+r_{K}-\frac{s}{2}\right) \left| F_{1}^{9}\left(
q^{2}\right) \right| ^{2}+\left( 1-r_{K}\right) \mathop{\rm Re}%
F_{1}^{9}\left( q^{2}\right) F_{2}^{9^{*}}\left( q^{2}\right) +\frac{s}{2}%
\left| F_{2}^{9}\left( q^{2}\right) \right| ^{2},  \label{deltak} \\
\beta _{K^{*}} &=&\left[ s\left( 2\varphi _{K^{*}}\tilde{V}\left(
q^{2}\right) +3\tilde{F}_{0}\left( q^{2}\right) \right) +\frac{\varphi
_{K^{*}}}{4r_{K^{*}}}\left( \tilde{F}_{0}\left( q^{2}\right) +\varphi
_{K^{*}}\tilde{F}_{1}\left( q^{2}\right) \right. \right.  \nonumber \\
&&\left. \left. +2\left( 1-r_{K^{*}}-s\right) \tilde{F}_{01}\left(
q^{2}\right) \right) \right] ,  \label{betak*} \\
\delta _{K^{*}} &=&\frac{\varphi _{K^{*}}}{2}\left[ -2\left| V^{9}\left(
q^{2}\right) M_{B}\right| ^{2}-\frac{3}{\varphi _{K^{*}}}\left| \frac{%
A_{0}^{9}\left( q^{2}\right) }{M_{B}}\right| ^{2}+\frac{s}{4r_{K^{*}}}\left|
A_{2}^{9}\left( q^{2}\right) M_{B}\right| ^{2}\right.  \nonumber \\
&&+\frac{2\left( 1+r_{K^{*}}\right) -s}{4r_{K^{*}}}\left| A_{1}^{9}\left(
q^{2}\right) M_{B}\right| ^{2}+\frac{1}{2r_{K^{*}}}\mathop{\rm Re}\left(
A_{0}^{9}\left( q^{2}\right) A_{1}^{9*}\left( q^{2}\right) +A_{0}^{9}\left(
q^{2}\right) A_{2}^{9*}\left( q^{2}\right) \right)  \nonumber \\
&&\left. +\frac{1-r_{K^{*}}}{2r_{K^{*}}}\mathop{\rm Re}A_{1}^{9}\left(
q^{2}\right) M_{B}A_{2}^{9*}\left( q^{2}\right) M_{B}\right] ,
\label{deltak*}
\end{eqnarray}
with
\begin{eqnarray}
\tilde{V}\left( q^{2}\right) &=&\left| V^{8}\left( q^{2}\right) M_{B}-\frac{%
2m_{b}M_{B}}{q^{2}}T^{7}\left( q^{2}\right) \right| ^{2}+\left| V^{9}\left(
q^{2}\right) M_{B}\right| ^{2},  \nonumber \\
\tilde{F}_{0}\left( q^{2}\right) &=&\left| \frac{A_{0}^{8}\left(
q^{2}\right) }{M_{B}}-\frac{2m_{b}}{q^{2}}\frac{T_{0}^{7}\left( q^{2}\right)
}{M_{B}}\right| ^{2}+\left| \frac{A_{0}^{9}\left( q^{2}\right) }{M_{B}}%
\right| ^{2},  \nonumber \\
\tilde{F}_{1}\left( q^{2}\right) &=&\left| A_{1}^{8}\left( q^{2}\right)
M_{B}-\frac{2m_{b}M_{B}}{q^{2}}T_{1}^{7}\left( q^{2}\right) \right|
^{2}+\left| A_{1}^{9}\left( q^{2}\right) M_{B}\right| ^{2},  \nonumber \\
\tilde{F}_{01}\left( q^{2}\right) &=&\mathop{\rm Re}\left[ \left( \frac{%
A_{0}^{8}\left( q^{2}\right) }{M_{B}}-\frac{2m_{b}}{q^{2}}\frac{%
T_{0}^{7}\left( q^{2}\right) }{M_{B}}\right) \left( A_{1}^{8*}\left(
q^{2}\right) M_{B}-\frac{2m_{b}M_{B}}{q^{2}}T_{1}^{7*}\left( q^{2}\right)
\right) \right]  \nonumber \\
&&+\mathop{\rm Re}\left( \frac{A_{0}^{9}\left( q^{2}\right) }{M_{B}}%
A_{1}^{9*}\left( q^{2}\right) M_{B}\right) .  \label{betaks}
\end{eqnarray}

The forward-backward asymmetry (FBA) can be defined by

\begin{equation}
{\cal A}_{FB}=\frac{1}{d\Gamma \left( s\right) /ds}\left[ \int_{0}^{1}d\cos
\theta \ \frac{d^{2}\Gamma \left( s\right) }{dsd\cos \theta }%
-\int_{-1}^{0}d\cos \theta \ \frac{d^{2}\Gamma \left( s\right) }{dsd\cos
\theta }\right] \,,
\end{equation}
where $\theta$ is the angle of charged $l^{+}$ with respect to the B meson
in the rest frame of the lepton pair. For $B\rightarrow K^{*}l^{+}l^{-}$
decay, the FBA is found to be

\begin{equation}
{\cal A}_{FB}^{K^{*}}=-\frac{3s\sqrt{\varphi _{K^{*}}}\sqrt{1-\frac{%
4m_{l}^{2}}{q^{2}}}R_{VA}\left( q^{2}\right) }{\left( 1+\frac{2m_{l}^{2}}{%
q^{2}}\right) \beta _{K^{*}}+12\frac{m_{l}^{2}}{M_{B}^{2}}\delta _{K^{*}}}
\label{fb}
\end{equation}
with
\begin{eqnarray}
R_{VA}\left( q^{2}\right) &=& \mathop{\rm Re} \left[ \left( V^{8}\left(
q^{2}\right) M_{B}-\frac{2m_{b}M_{B}}{q^{2}}T^{7}\left( q^{2}\right) \right)
\frac{A_{0}^{9*}\left( q^{2}\right) }{M_{B}}\right]  \nonumber \\
&&+\mathop{\rm Re}\left[ \left( \frac{A_{0}^{8}\left( q^{2}\right) }{M_{B}}-%
\frac{2m_{b}}{q^{2}}\frac{T_{0}^{7}\left( q^{2}\right) }{M_{B}}\right)
V^{9*}\left( q^{2}\right) M_{B}\right] \,.
\end{eqnarray}
As expected, the FBA in Eq. (\ref{fb}) is sensitive to the chiral structure
of interactions since it is related to the product of $V$ and $A$ currents.
It is clear that the FBA for $B\rightarrow Kl^{+}l^{-}$ vanishes since there
is no form factor from the axial current.

Another interesting lepton asymmetry is the longitudinal polarization of the
lepton, defined by

\[
P_{L}\left( q^{2}\right) =\frac{\frac{d\Gamma \left( n=-1\right) }{ds}-\frac{%
d\Gamma \left( n=1\right) }{ds}}{\frac{d\Gamma \left( n=-1\right) }{ds}+%
\frac{d\Gamma \left( n=1\right) }{ds}}
\]
where $n$ is the projection of the lepton $l^{-}$ momentum to the spin
direction in its rest frame. For $B\rightarrow \left( K,K^{*}\right)
l^{+}l^{-}$, the polarization asymmetries can be expressed as

\begin{equation}
P_{L}^{K}\left( q^{2}\right) =\frac{2\sqrt{1-\frac{4m_{l}^{2}}{q^{2}}}%
\varphi _{K}}{\left( 1+\frac{2m_{l}^{2}}{q^{2}}\right) \beta _{K}+12\frac{%
m_{l}^{2}}{M_{B}^{2}}\delta _{K}}\mathop{\rm Re}\left[ \left(
F_{1}^{8}\left( q^{2}\right) -2m_{b}F_{T}^{7}\left( q^{2}\right) \right)
F_{1}^{9*}\left( q^{2}\right) \right]  \label{plk}
\end{equation}
and

\begin{eqnarray}
P_{L}^{K^{*}}\left( q^{2}\right) &=&\frac{2\sqrt{1-\frac{4m_{l}^{2}}{q^{2}}}%
}{\left( 1+\frac{2m_{l}^{2}}{q^{2}}\right) \beta _{K^{*}}+12\frac{m_{l}^{2}}{%
M_{B}^{2}}\delta _{K^{*}}}\left\{ s\left[ 2\varphi _{K^{*}}R_{V}\left(
q^{2}\right) +3R_{A_{0}}\left( q^{2}\right) \right] \right.  \nonumber \\
&&\left. +\frac{\varphi _{K^{*}}}{4r}\left[ R_{A_{0}}\left( q^{2}\right)
+\varphi _{K^{*}}R_{A_{1}}\left( q^{2}\right) +\left( 1-r_{K^{*}}-s\right)
R_{A_{01}}\left( q^{2}\right) \right] \right\}  \label{plk*}
\end{eqnarray}
with

\begin{eqnarray}
R_{V}\left( q^{2}\right) &=&\mathop{\rm Re}\left[ \left( V^{8}\left(
q^{2}\right) M_{B}-\frac{2m_{b}M_{B}}{q^{2}}T^{7}\left( q^{2}\right) \right)
V^{9*}\left( q^{2}\right) M_{B}\right] ,  \nonumber \\
R_{A_{0}}\left( q^{2}\right) &=&\mathop{\rm Re}\left[ \left( \frac{%
A_{0}^{8}\left( q^{2}\right) }{M_{B}}-\frac{2m_{b}}{q^{2}}\frac{%
T_{0}^{7}\left( q^{2}\right) }{M_{B}}\right) \frac{A_{0}^{9*}\left(
q^{2}\right) }{M_{B}}\right] ,  \nonumber \\
R_{A_{1}}\left( q^{2}\right) &=&\mathop{\rm Re}\left[ \left( A_{1}^{8}\left(
q^{2}\right) M_{B}-\frac{2m_{b}M_{B}}{q^{2}}T_{1}^{7}\left( q^{2}\right)
\right) A_{1}^{9*}\left( q^{2}\right) M_{B}\right] ,  \nonumber \\
R_{A_{01}}\left( q^{2}\right) &=&\mathop{\rm Re}\left[ \left( \frac{%
A_{0}^{8}\left( q^{2}\right) }{M_{B}}-\frac{2m_{b}}{q^{2}}\frac{%
T_{0}^{7}\left( q^{2}\right) }{M_{B}}\right) A_{1}^{9*}\left( q^{2}\right)
M_{B}\right.  \nonumber \\
&&\left. +\left( A_{1}^{8}\left( q^{2}\right) M_{B}-\frac{2m_{b}M_{B}}{q^{2}}%
T_{1}^{7}\left( q^{2}\right) \right) \frac{A_{0}^{9*}\left( q^{2}\right) }{%
M_{B}}\right] ,
\end{eqnarray}
respectively.

\section{Numerical Analysis}

\subsection{Form factors}

In Eq. (\ref{FT}), $\phi (x,b)$ is the universal wave function and cannot be
calculated perturbatively. However, due to the universality, we can
determine it by matching with the $B$ decay experimental data. With the
ratio of
\begin{equation}
R=\frac{B(B_{d}^{0}\to K^{\pm }\pi ^{\mp })}{B(B^{\pm }\to K^{0}\pi ^{\pm })}%
=0.95\pm 0.3\;,
\end{equation}
given by the CLEO measurement \cite{CLEO3}, where $B(B_{d}^{0}\to K^{\pm
}\pi ^{\mp })$ represents the CP average of the branching ratios $%
B(B_{d}^{0}\to K^{+}\pi ^{-})$ and $B({\bar{B}}_{d}^{0}\to K^{-}\pi ^{+})$,
one can get the proper wave functions $\phi _{B}$, $\phi _{K}$, and $\phi
_{K}^{\prime }$ \cite{KLS} while $\phi _{K^{*}}$ can be done by the
branching ratio of $B\rightarrow K^{*}\gamma $ \cite{LL}. For the $B$ meson
wave function, we take
\begin{equation}
\phi _{B}(x,b)=N_{B}x^{2}(1-x)^{2}\exp \left[ -\frac{1}{2}\left( \frac{xM_{B}%
}{\omega _{B}}\right) ^{2}-\frac{\omega _{B}^{2}b^{2}}{2}\right] ,
\label{bwf}
\end{equation}
with the shape parameter $\omega _{B}=0.4$ GeV \cite{BW}. The normalization
constant $N_{B}=91.7835$ GeV is related to the decay constant $f_{B}=190$
MeV. The kaon wave functions are chosen as
\begin{eqnarray}
\phi _{K}(x) &=&\frac{3}{\sqrt{2N_{c}}}%
f_{K}x(1-x)[1+0.51(1-2x)+0.3(5(1-2x)^{2}-1)]\;,  \nonumber \\
\phi _{K}^{\prime }(x) &=&\frac{3}{\sqrt{2N_{c}}}f_{K}x(1-x),  \nonumber \\
\phi _{K^{*}}(x) &=&\frac{3}{\sqrt{2N_{c}}}%
f_{K^{*}}x(1-x)[1+0.51(1-2x)+(5(1-2x)^{2}-1)]\,,  \label{wf}
\end{eqnarray}
where $\phi _{K}$ is derived from QCD sum rules \cite{PB2}, and the second
term in the expression of $\phi _{K}$ corresponds to $SU(3)$ symmetry
breaking effect. The decay constants $f_{K}$ and $f_{K^{*}}$ are set to be $%
160$ and $190$ MeV (in the convention of $f_{\pi }=130$ MeV),
respectively. Note that the intrinsic $b$ dependences of wave
functions in Eq. (\ref{wf}) are neglected. However, this is a good
approximation only for the fast recoiling meson, that is, the
transverse extent of the wave function is less important in the
energetic outgoing situation. With above wave functions and
taking $M_{B}=5.28$ GeV, $M_{K}=0.49$ GeV, $m_{s}=100$ MeV and $%
M_{K^{*}}=0.89$ GeV, the form factors of $B\rightarrow K$ defined in Eqs. (%
\ref{ff}) as a function of $q^{2}$ are shown in Figure 1. The
values of the form factors at $q^{2}=0$ are given in Table 1. We
now compare our results with that in the light cone-QCD sum rule
(LCSR ) \cite{Bmeson}. Using the identities in the Appendix, we
find that except $V(0)$ is slightly smaller than that of the
minimal value, while the remaining form factors are within the
allowed values, in the LCSR.
\begin{table}[h]
\caption{Form factors at $q^{2}=0$ in the PQCD.}
\begin{center}
\begin{tabular}{|cccccccccc|}
\hline
$F_{1}(0)$ & $F_{2}(0)$ & $F_{T}(0)$ & $V(0)$ & $A_{0}(0)$ & $A_{1}(0)$ & $%
A_{2}(0)$ & $T(0)$ & $T_{1}(0)$ & $T_{2}(0)$ \\ \hline\hline
$0.33$ & $-0.267$ & $-0.054$ & $0.063$ & $2.02$ & $-0.05$ & $0.059$ & $%
-0.350 $ & $0.350$ & $-0.281$ \\ \hline
\end{tabular}
\end{center}
\end{table}
Recently, it has been mentioned that by combining large energy
effective theory (LEET) \cite{Charles}, originally proposed by
\cite{DG}, with the
measurement of $B\rightarrow K^{*}\gamma $, the form factors $V(0)$ and $%
A_{0}(0)$ could be fitted model-independently to be \cite{BH}
\begin{eqnarray}
V(0) &\simeq &0.069\pm 0.011,  \nonumber \\
A_{0}(0) &\simeq &1.650\pm 0.114.  \label{fv}
\end{eqnarray}
Hence, from Table 1, we clearly see that the values of $V(0)$ and $A_{0}(0)$
are within $1\sigma $ and $3.2\sigma $ of values in Eq. (\ref{fv}),
respectively. It is worth to mention that the LEET predicts $T_{2}(0)/A_{1}(0)%
\sim 4.89$ \cite{BH}, while that in our approach is $5.62$.
 For the comparison of the form factors between different models at $%
q^{2}=0$, one can refer to Ref. \cite{BH} for a more detail
analysis.

For the exclusive $B\rightarrow K^{(*)}l^{+}l^{-}$ decays, if we use $b$
independent wave functions, as the mesons reach the slow recoil, the
suppression in the large $b$ region is weaker such that form factors will
blow up at points away from $q^{2}=0$ as seen from Figure 1. It is
inevitable to include the intrinsic $b$ dependence to the outgoing meson
wave functions. However, the effect of such dependence is less significant
for the small $q^{2}$ values than that of large ones. Instead of using an
exponential $b$ dependent form as the one for the B meson wave function in
Eq. (\ref{bwf}), for simplicity, we use the trial wave functions as %

\begin{equation}
\phi _{H}^{\left( \prime \right) }(x,q^{2})=(1-\frac{q^{2}}{M_{B}^{2}})\sqrt{%
1+\frac{q^{2}}{M_{B}^{2}}}\phi _{H}^{\left( \prime \right) }\left( x\right) .
\label{mwf}
\end{equation}
On the other hand, since the available region of the PQCD essentially cannot
include all allowed values of $q^{2}$, to have the form factors in the whole
accessible values of $q^{2}$, we would adopt the parametrization of
effective form factors as follows:

\begin{equation}
F\left( s\right) =F\left( 0\right) \exp \left( \sigma _{1}s+\sigma
_{2}s^{2}+\sigma _{3}s^{3}\right)  \label{exp}
\end{equation}
with $s=q^{2}/M_{B}^{2}$ to fit the values up to $q^{2}\approx 15$ GeV$^{2}$
calculated by the PQCD. By extrapolating to near the endpoint of $q^{2}$, we
found that the values of form
factors are consistent with lattice results \cite{APE,UKQCD}. To illustrate
how good the trial functions in Eq. (\ref{mwf}) are, we show the form
factors for $B\to K$ in Figure 2. From the figure, we find that our results
are basically the same as that from the QM \cite{MNS} and LCSR \cite{Bmeson}.


\subsection{Decay rates}

With the confidence of calculating the form factors by using Eqs. (\ref{mwf}%
) and (\ref{exp}), we now study the decay rates of $B\to K^{(*)}l^{+}l^{-}$.
Unlike the conventional FA, the WCs in Eqs. $\left( \ref{ampk}\right) $ and $%
\left( \ref{ampk*}\right) $ are the members of integrations in the PQCD.
Thus, adopting the approach similar to form factors, we calculate the
transition amplitudes with the wave functions in Eq. (\ref{mwf}) and the
exponential forms in Eq. (\ref{exp}) to fit the values calculated by the
PQCD for the whole range of $q^2$. After integrating $q^{2}$ dependence in
Eq. (\ref{dr}), the decay branching ratios without including LD
contributions for 
$B\rightarrow \left( K,K^{*}\right) l^{+}l^{-}$ are listed in Table \ref{br}
and their distributions for the differential decay rates are shown in Figure
3. Comparing with the curves in the QM and LCSR, we find that the
differential rates of $B\rightarrow Kl^{+}l^{-}$ are consistent with each
other. However, there exists a slight difference in $B\rightarrow
K^{*}l^{+}l^{-}$. There are two main reasons for the difference: (a) $\mu $
scale dependence for the WC and (b) the effects from $A_{1}(q^{2})$ and $%
A_{2}(q^{2})$ in the PQCD. For the results in QM and LCSR, we have used the
WC at $\mu \sim m_{b}$ as done in the literature, whereas that in the PQCD, $%
\mu $ scale is the typical scale $t$ determined by Eq. (\ref{tscale}). As
seen from Eq. (\ref{betaks}) the effect of $A_{1}(q^{2})$ is large since
there is a factor of $M_{B}$ associated with it, while that of $A_{2}(q^{2})$
only affects in the mode of $B\rightarrow K^{*}\tau ^{+}\tau ^{-}$ due to
the lepton mass dependence. Thus, measuring the exclusive modes of $%
B\rightarrow K^{*}l^{+}l^{-}$ would distinguish various QCD models due to
the difference shown in Figure 3. From Table \ref{br}, we see that our PQCD
results of the decay branching ratios for $B\rightarrow K^{*}\mu ^{+}\mu
^{-} $ and $B\rightarrow K^{*}e^{+}e^{-}$ are quite different. This can be
understood by noting that in Eq. (\ref{betak*}) there is pole of $q^{2}$
associated with the photon penguin induced couplings. These pole terms make
the rates sensitive to the kinematical region of $q^{2}\geq 4m_{l}^{2}$.

\begin{table}[h]
\caption{Deacy branching ratios in the various QCD models without including
LD effects.}
\label{br}
\begin{center}
\begin{tabular}{|c|ccc|}
\hline
Mode & PQCD & QM & LCSR \\ \hline
$10^{7}$B$(B\rightarrow Kl^{+}l^{-}) $ & $5.33$ & $5.56$ & $5.20$ \\
$10^{7}$B$( B\rightarrow K\tau ^{+}\tau ^{-}) $ & $1.29$ & $1.28$ & $1.25$
\\ \hline
$10^{6}$B$( B\rightarrow K^{*}e^{+}e^{-}) $ & $2.26$ & $1.88$ & $2.23$ \\
$10^{6}$B$( B\rightarrow K^{*}\mu ^{+}\mu ^{-}) $ & $1.27$ & $1.49$ & $1.78$
\\
$10^{7}$B$( B\rightarrow K^{*}\tau ^{+}\tau ^{-}) $ & $1.24$ & $1.43$ & $%
1.77 $ \\ \hline
\end{tabular}
\end{center}
\end{table}

\subsection{Forward-backward asymmetry}

>From Eq. (\ref{fb}), we present the forward-backward dilepton asymmetries of
$B\to K^{*}l^{+}l^{-}\ (l=\mu,\tau)$ in Figure 4. We note that the FBA of $%
B\to K^{*}e^{+}e^{-}$ is similar to that of the muon mode in Figure 4. For
the light lepton pair decays such as $B\to K^{*}\mu^{+}\mu^{-}$, we see that
the FBA is positive at low $q^2$, gets zero at $q^{2}/M_{B}^{2}\sim 0.16$,
and then becomes negative. Due to the large terms related to $A_{1}(q^{2})$,
the rate at lower $q^2$ in the PQCD has a larger value than that in the QM
and LCSR. As mentioned in Ref. \cite{Bmeson} with the FA, the location of
zero point is only sensitive to the WC and insensitive to the form factors.
However, since with the three-scale factorization formula the WC cannot be
factored out of the transition amplitude and it is uncertain to choose the
universal wave function of $K^{*} $, the determination of zero point is
harder as the test of the SM in the approach of the PQCD, unless we can fix $%
K^{*}$ wave function more precisely. On the other hand, it is worth to
mention that when the $sign( C_8C_7)=+$, opposite to the SM, the zero point
disappears. Thus the FBA is quite sensitive to the sign of the WC and can be
used as a good candidate to test the SM.

\subsection{Polarization asymmetry}

The lepton polarization asymmetries of $B\rightarrow \left( K,K^{*}\right)
l^{+}l^{-}$ are displayed in Figure 5. For $B\rightarrow Kl^{+}l^{-}$, $P_L$
is equal zero at $q^{2}=0$ and $q^{2}|_{\max }=\left( M_{B}-M_{K}\right)
^{2} $ because it is related to $\sqrt{1-4m_{l}^{2}/q^{2}}\varphi _{K}$.
Without LD effects for the light leptons $(l=e,\mu )$, from Eqs. (\ref{betak}%
) and (\ref{plk}), we easily realize that $P_{L}$ is near $-1$ in the most
region of $q^{2}$. However, for $B\rightarrow K^{*}l^{+}l^{-} $, since $%
\varphi _{K^{*}}$ cannot be factored out in the numerator, there is no
vanishing point at $q^{2}|_{\max }$; and the transition matrix elements of
the set $\left\{ T^{7}\right\} $ are always associated with a pole of $\
q^{2}$. Hence, at the low momentum transfer region, penguin induced
electromagnetic effects are dominant. Also comparing Eq. (\ref{plk*}) with
Eq. (\ref{betak*}), the related terms of $\left\{ T^{7}\right\} $ are two
powers for the differential decay rate but only one power for the numerator
of $P_{L}$ so that the magnitude of the distribution at low $q^{2}$ has a
smaller value. From Figures 5c and 5d, the polarization asymmetry of $%
B\rightarrow K^{*}\mu ^{+}\mu ^{-}$ in the PQCD has slightly different
distribution to other models at low $q^{2}$, especially that the deviation
for $B\rightarrow K^{*}\tau ^{+}\tau ^{-}$ is large. However, according to
Figure 6, we find that our results are comparable with that given by the
light-front formalism (LF) \cite{GIW,GK}.
As mentioned before, the influence of both larger values from the $%
A_{1}\left( q^{2}\right) $ and $A_{2}\left( q^{2}\right) $ terms in our
approach is visible for the $\tau $ mode. Therefore, by measuring the
longitudinal $\tau$ polarization in $B\rightarrow K^{*}\tau ^{+}\tau ^{-}$,
we can either determine a more proper $K^{*}$ wave function or test the
feasibility of our PQCD approach for semileptonic decays.

\section{Conclusions}

By the three-scale factorization theorem, we have gotten the form factors in
$q^{2}\leq 15$ GeV$^{2}$; and with the parametrization in terms of the
exponential forms to extrapolate the $q^2$ dependent form factors to $%
q_{\max}^{2}$, we have obtained the consistent results with that from the
lattice \cite{APE,UKQCD}. With the PQCD, we have pointed out that the
largest uncertainty in our results is from the non-perturbative wave
functions. Though the universal wave function could be determined by some
non-leptonic decays, the intrinsic $b$ dependence which suppresses the soft
dynamics contribution is still unknown.

With the kaon wave functions fixed by the decays of $B\rightarrow \pi K$ and
assumed $q^{2}$ dependences, we have shown that the distributions of the
decay rates and leptonic asymmetries for $B\rightarrow Kl^{+}l^{-}$ in the
PQCD agree well with that in the other QCD models such as the QM and LCSR.
However, there are some differences for that in $B\rightarrow
K^{*}l^{+}l^{-} $ among the various QCD models. Finally, we remark that
although $B\rightarrow K^{*}\gamma $ could also give us some information of
the $K^{*}$ wave function, one still cannot fix it satisfactorily due to
various uncertainties in the decay. Moreover, the assumption of the same $%
q^{2}$ dependent factors in $B\rightarrow K$ is not necessary for $%
B\rightarrow K^{*}$ and thus, to have a reliable calculation, we need more
precision measurements involving the vector kaon meson in order to settle
down the $K^{*}$ wave function. \newline

{\bf Acknowledgments}

We would like to thank Hsiang-nan Li for useful discussions. This work was
supported in part by the National Science Council of the Republic of China
under contract numbers NSC-89-2112-M-007-054 and NSC-89-2112-M-006-033.

\newpage

\bc
Appendix
\ec


In order to connect our form factors in Eq. (\ref{ff}) to those
usually used in the literature \cite{Bmeson,MNS,BH,GIW}, in this
Appendix we show explicitly the relationships among the form
factors. In terms of the notation in Refs. \cite{Bmeson,BH}, the
form factors for $B\rightarrow \left( K,K^{*}\right) $ decays with
respect to various weak currents are parametrized as

\begin{eqnarray}
\left\langle K\left( P_{2}\right) |\ V_{\mu }\ |B\left( P_{1}\right)
\right\rangle &=&f_{+}\left( q^{2}\right) \left( P_{\mu }-\frac{P\cdot q}{%
q^{2}}q_{\mu }\right)  \nonumber \\
&&+\frac{P\cdot q}{q^{2}}f_{0}\left( q^{2}\right) q_{\mu },  \label{bp1} \\
\left\langle K\left( P_{2}\right) |\ T_{\mu \nu }q^{\nu }\ |B\left(
P_{1}\right) \right\rangle &=&-\left( P_{\mu }q^{2}-q_{\mu }P\cdot q\right)
f_{T}\left( q^{2}\right) ,  \label{bp2}
\end{eqnarray}
\begin{eqnarray}
\left\langle K^{*}\left( P_{2},\varepsilon \right) |\ V_{\mu }\ |B\left(
P_{1}\right) \right\rangle &=&-\frac{V^{\prime }\left( q^{2}\right) }{%
M_{B}+M_{K^{*}}}\epsilon _{\mu \nu \alpha \beta }\varepsilon ^{*\nu
}P^{\alpha }q^{\beta },  \label{bv1} \\
\left\langle K^{*}\left( P_{2},\varepsilon \right) |\ A_{\mu }\ |B\left(
P_{1}\right) \right\rangle &=&i2M_{K^{*}}A_{0}^{\prime }\left( q^{2}\right)
\frac{\varepsilon ^{*}\cdot q}{q^{2}}q_{\mu }  \nonumber \\
&&+i\left( M_{B}+M_{K^{*}}\right) A_{1}^{\prime }\left( q^{2}\right) \left(
\varepsilon ^{*\mu }-\frac{\varepsilon ^{*}\cdot q}{q^{2}}q_{\mu }\right)
\nonumber \\
&&-iA_{2}^{\prime }\left( q^{2}\right) \frac{\varepsilon ^{*}\cdot q}{%
M_{B}+M_{K^{*}}}\left( P_{\mu }-\frac{P\cdot q}{q^{2}}q_{\mu }\right) ,
\label{bv2} \\
\left\langle K^{*}\left( P_{2},\varepsilon \right) |\ T_{\mu \nu }q^{\nu }\
|B\left( P_{1}\right) \right\rangle &=&T_{1}^{\prime }\left( q^{2}\right)
\epsilon _{\mu \nu \alpha \beta }\varepsilon ^{*\nu }P^{\alpha }q^{\beta },
\label{bv3} \\
\left\langle K^{*}\left( P_{2},\varepsilon \right) |\ T_{\mu \nu }^{5}q^{\nu
}\ |B\left( P_{1}\right) \right\rangle &=&iT_{2}^{\prime }\left(
q^{2}\right) \left[ \varepsilon _{\mu }^{*}P\cdot q+\varepsilon ^{*}\cdot
qP_{\mu }\right]  \nonumber \\
&&-iT_{3}^{\prime }\left( q^{2}\right) \varepsilon ^{*}\cdot q\left[ q_{\mu
}-\frac{q^{2}}{P\cdot q}P_{\mu }\right] ,  \label{bv4}
\end{eqnarray}
where $V_{\mu }=\ \bar{s}\ \gamma _{\mu }\ b$, $A_{\mu }=\bar{s}\ \gamma
_{\mu }\gamma _{5}\ b$, $T_{\mu \nu }=\bar{s}\ i\sigma _{\mu \nu }b$, $%
T_{\mu \nu }^{5}=\bar{s}\ i\sigma _{\mu \nu }\gamma _{5}\ b,$ $P=P_{1}+P_{2}$%
, $q=P_{1}-P_{2}$, and $P\cdot q=M_{B}^{2}-M_{K^{(*)}}^{2}$.
Redefining the wave functions and comparing to Eq. (\ref{ff}), we
obtain
\begin{eqnarray*}
F_{1} &=&f_{+} \\
F_{2} &=&\frac{M_{B}^{2}-M_{K}^{2}}{q^{2}}\left( f_{0}-f_{+}\right) \\
V &=&\frac{V^{\prime }}{M_{B}+M_{K^{*}}} \\
A_{0} &=&\left( M_{B}+M_{K^{*}}\right) A_{1}^{\prime } \\
A_{1} &=&-\frac{A_{2}^{\prime }}{M_{B}+M_{K^{*}}} \\
A_{2} &=&\frac{1}{q^{2}}\left[ 2M_{K^{*}}A_{0}^{\prime }-\left(
M_{B}+M_{K^{*}}\right) A_{1}^{\prime }+\left( M_{B}-M_{K^{*}}\right)
A_{2}^{\prime }\right] \\
T &=&-T_{1}^{\prime } \\
T_{0} &=&-\left( M_{B}^{2}-M_{K^{*}}^{2}\right) T_{2}^{\prime } \\
T_{1} &=&T_{2}^{\prime }+\frac{q^{2}}{M_{B}^{2}-M_{K^{*}}^{2}}T_{3}^{\prime }
\\
T_{2} &=&-T_{3}^{\prime }\,.
\end{eqnarray*}
Here we have neglected to show the $q^{2}$ dependence for the form
factors. From the above identities, we find some interesting
relations at $q^{2}=0$ and they are given by
\begin{eqnarray*}
f_{0}(0) &=&f_{+}(0), \\
T_{1}(0) &=&T_{2}^{\prime }(0).
\end{eqnarray*}
 From Eqs. (\ref{t}) and (\ref{t1}), we get $T(0)=-T_{1}(0).$ Hence, based
on the modified PQCD factorization theorem, we obtain the relation $%
T_{1}^{\prime }\left( 0\right) =T_{2}^{\prime }(0)$ that is the
same as that in Eq. (3.6) of Ref. \cite{Bmeson}.

\newpage
\begin{figcap}


\item
Form factors of $F_1$ (solid curves), $F_2$ (dashed curves)
and $F_T$ (dotted curves) for the $B\to K$ transition as a
function $q^2$ in the PQCD (bold lines) and QM (un-bold lines),
respectively.

\item
Form factors of $F_1$, $F_2$ and $F_T$ for the $B\to K$ transition as a
function $q^2$  with the $q^2$ dependent wave function in Eq. (\ref{mwf})
in the PQCD (solid curves), QM (dot-dashed curves), and LCSR (dashed
curves), respectively.

\item
The differential decay braching ratios as function of $s$
for (a) $B\rightarrow K\mu ^{+}\mu ^{-}$ and (b)
$B\rightarrow K\tau ^{+}\tau ^{-}$.
The curves with and without resonant shapes represent including and no
LD contributions, respectively.
Legend is the same as Figure 2.

\item
The differential decay braching ratios as function of $s$
for
(a) $B\rightarrow K^{*}\mu ^{+}\mu ^{-}$ and (b)
$B\rightarrow K^{*}\tau ^{+}\tau ^{-}$.
Legend is the same as Figure 3.

\item
 Forward-backward asymmetries for
(a) $B\rightarrow K^{*}\mu ^{+}\mu ^{-}$ and (b)
$B\rightarrow K^{*}\tau ^{+}\tau ^{-}$.
Legend is the same as Figure 3.

\item
Longitudinal polarization asymmetries
for (a) $B\rightarrow K\mu ^{+}\mu ^{-}$ and (b)
$B\rightarrow K\tau ^{+}\tau ^{-}$.
Legend is the same as Figure 3.

\item
Longitudinal polarization asymmetries
(a) $B\rightarrow K^{*}\mu ^{+}\mu ^{-}$ and (b)
$B\rightarrow K^{*}\tau ^{+}\tau ^{-}$.
Legend is the same as Figure 3.

\item
The longitudinal polarization asymmetry for
$B\rightarrow K^{*}\tau ^{+}\tau ^{-}$ in
the PQCD (solid curves) and LF (dashed curves), respectively.

\end{figcap}

\newpage
\begin{figure}[h]
\includegraphics{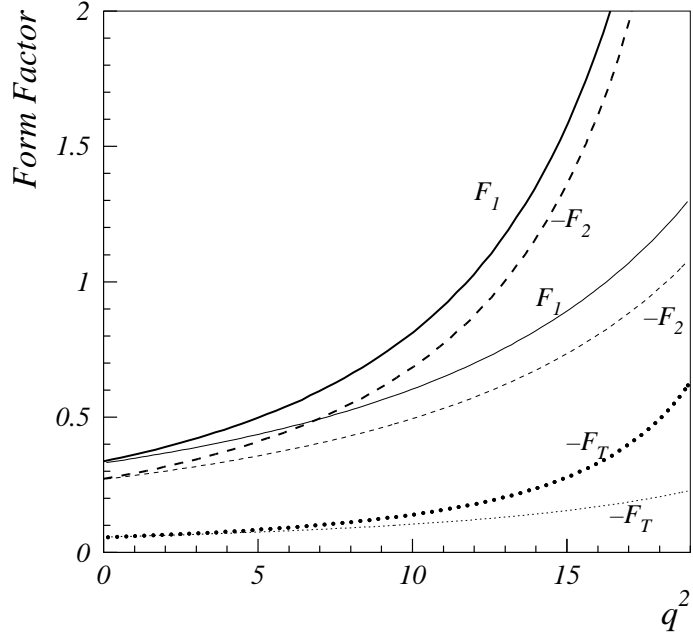} %
\vskip 13cm
\caption{ Form factors of $F_1$ (solid curves), $F_2$ (dashed curves) and $%
F_T$ (dotted curves) for the $B\to K$ transition as a function $q^2$ in the
PQCD (bold lines) and QM (un-bold lines), respectively. }
\end{figure}

\newpage
\begin{figure}[h]
\includegraphics{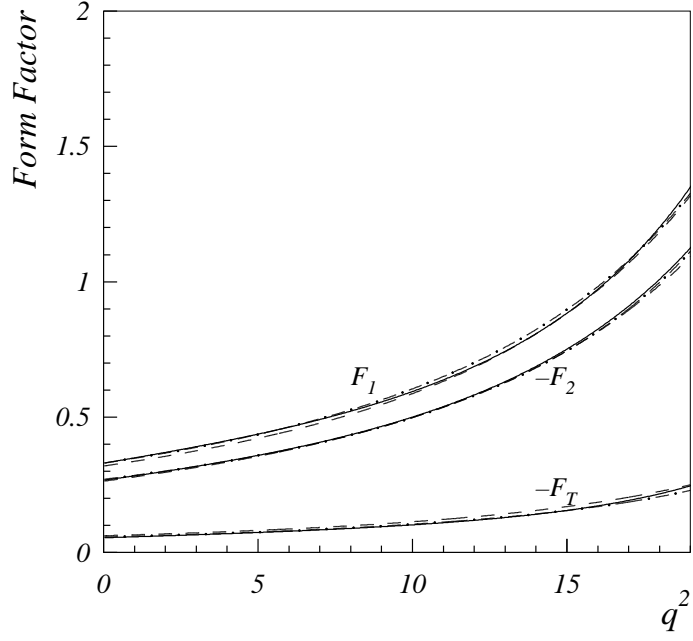} %
\vskip 13cm
\caption{ Form factors of $F_1$, $F_2$ and $F_T$ for the $B\to K$ transition
as a function $q^2$ with the $q^2$ dependent wave function in Eq. (\ref{mwf}%
) in the PQCD (solid curves), QM (dot-dashed curves), and LCSR (dashed
curves), respectively. }
\end{figure}

\newpage
\begin{figure}[h]
\includegraphics{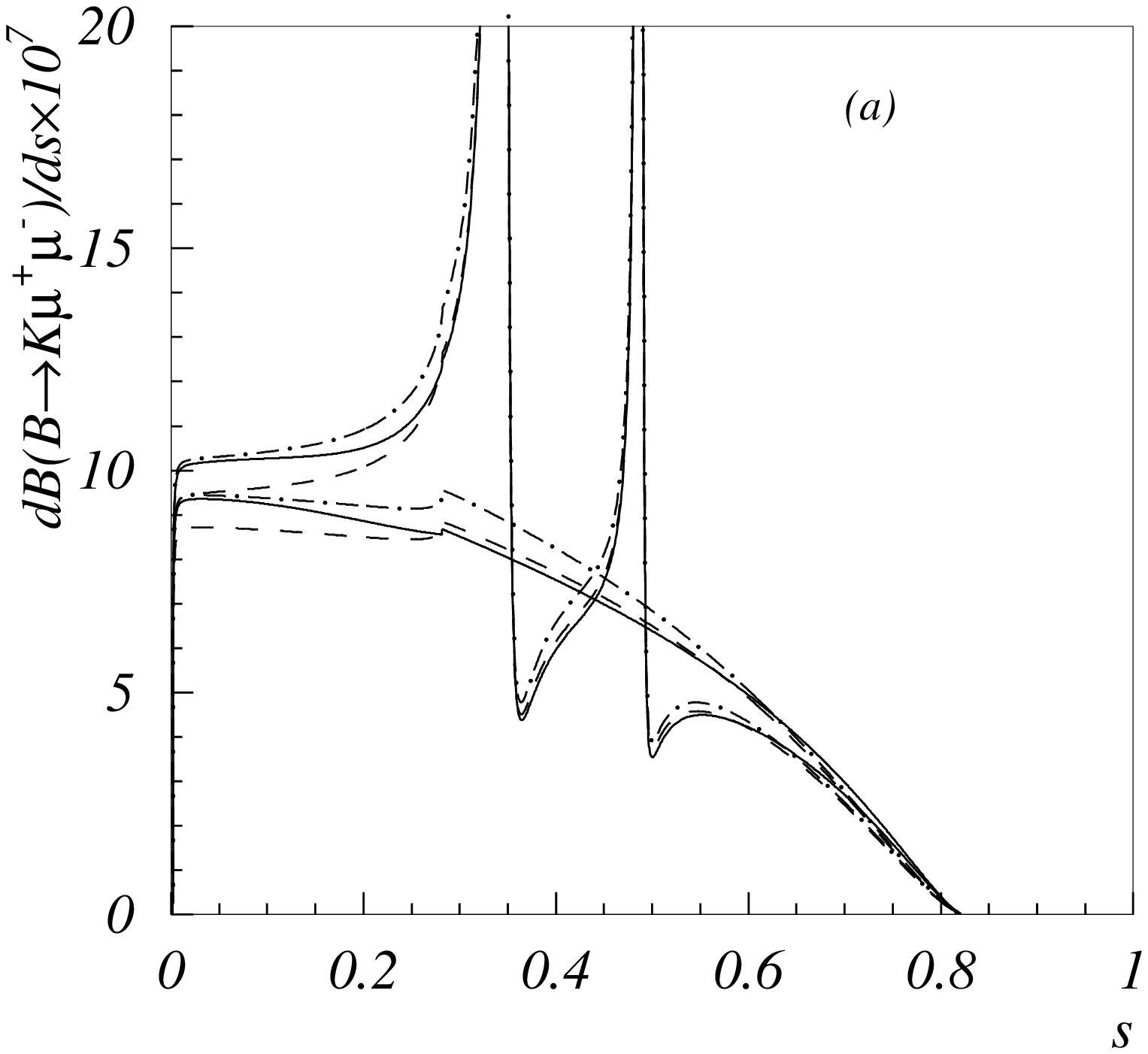}
\vskip 5.5cm
\end{figure}

\vskip 2.cm
\begin{figure}[h]
\includegraphics{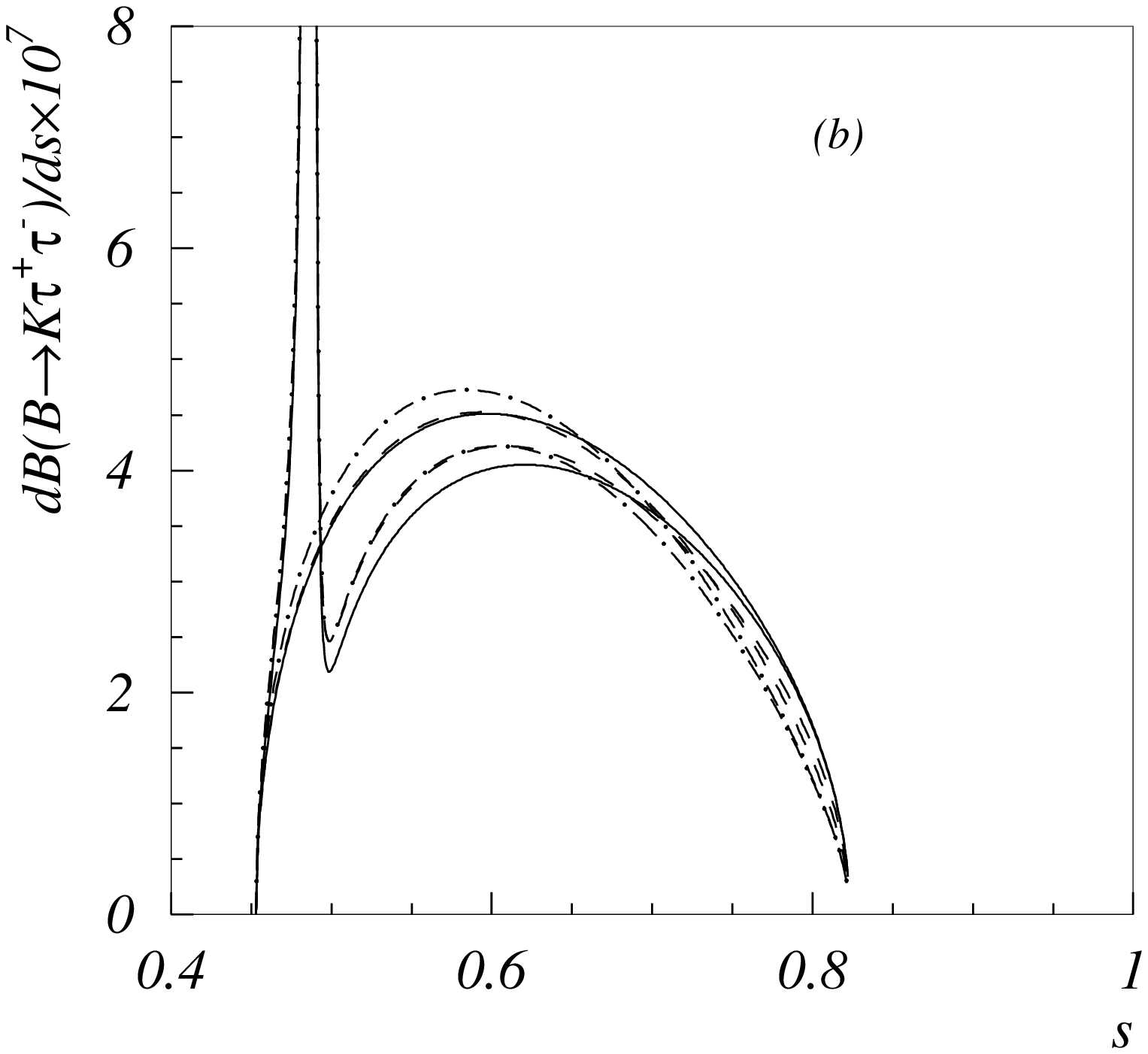}
\vskip 8.cm
\caption{ The differential decay braching ratios as function of $s$ for (a) $%
B\rightarrow K\mu ^{+}\mu ^{-}$ and (b) $B\rightarrow K\tau ^{+}\tau ^{-}$.
The curves with and without resonant shapes represent including and no LD
contributions, respectively. Legend is the same as Figure 2. }
\end{figure}

\newpage
\begin{figure}[h]
\includegraphics{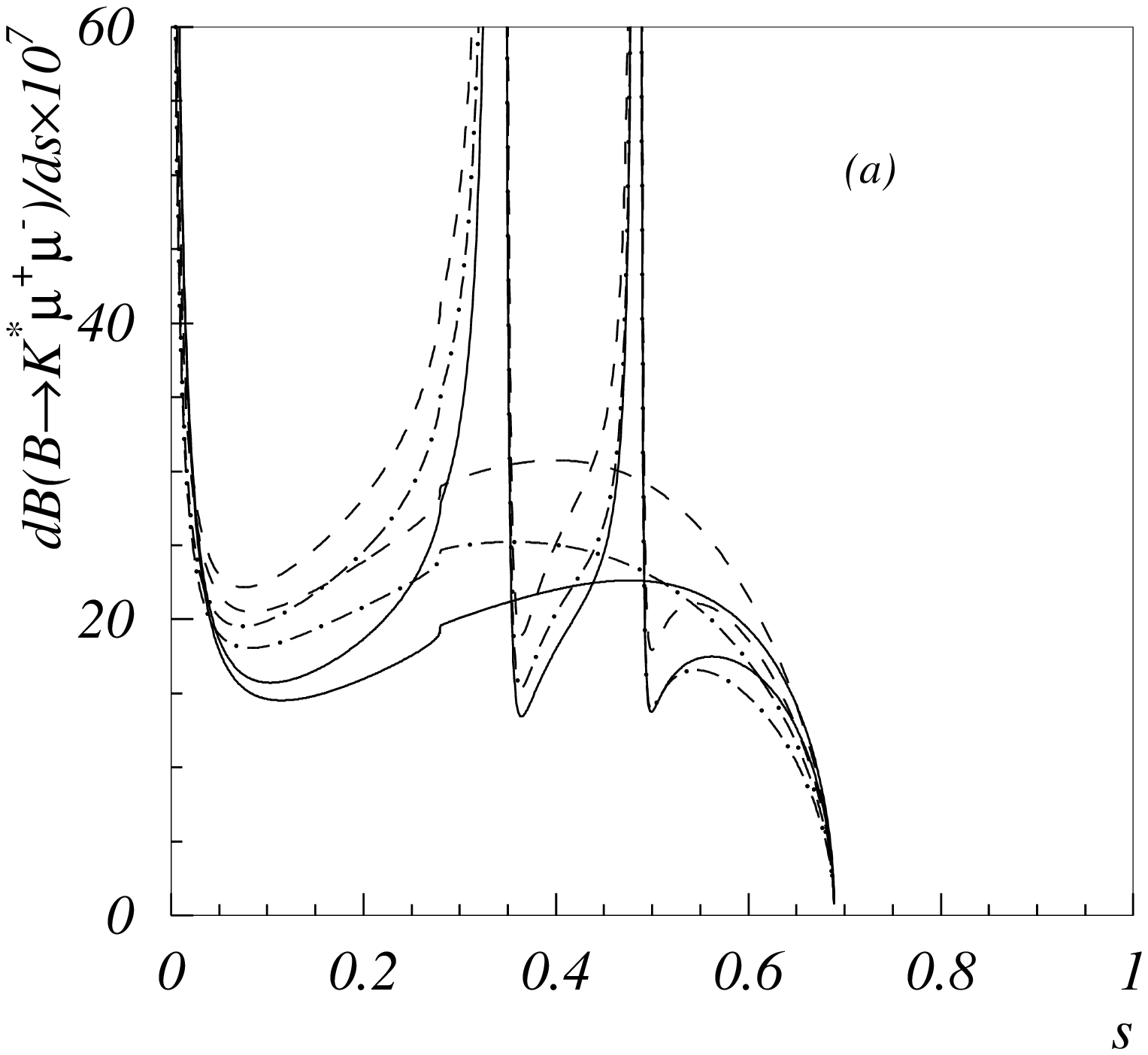}
\vskip 5.5cm
\end{figure}

\vskip 2.cm
\begin{figure}[h]
\includegraphics{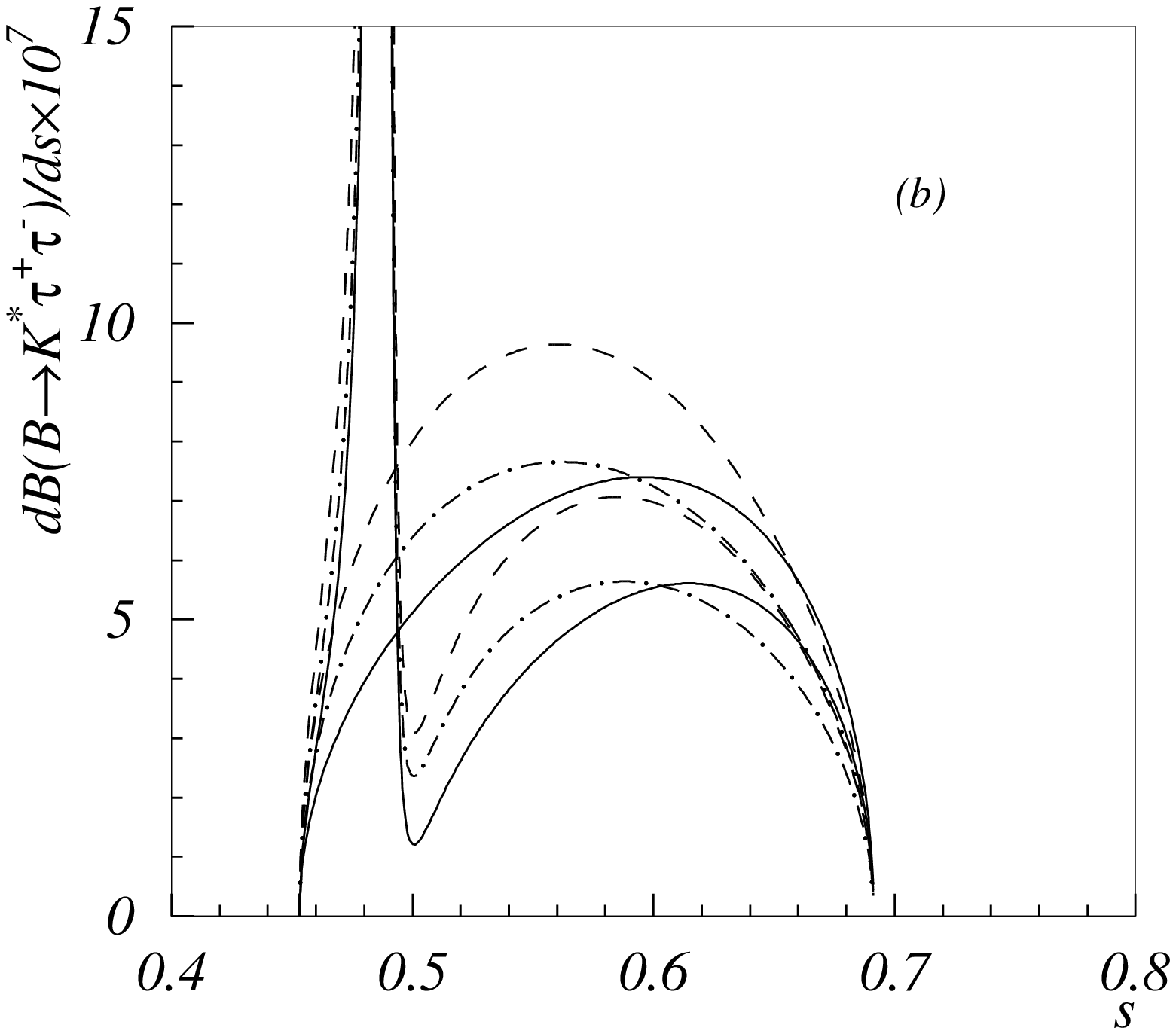}
\vskip 8.cm
\caption{ The differential decay braching ratios as function of $s$ for (a) $%
B\rightarrow K^{*}\mu ^{+}\mu ^{-}$ and (b) $B\rightarrow K^{*}\tau ^{+}\tau
^{-}$. Legend is the same as Figure 3. }
\end{figure}

\newpage
\begin{figure}[h]
\includegraphics{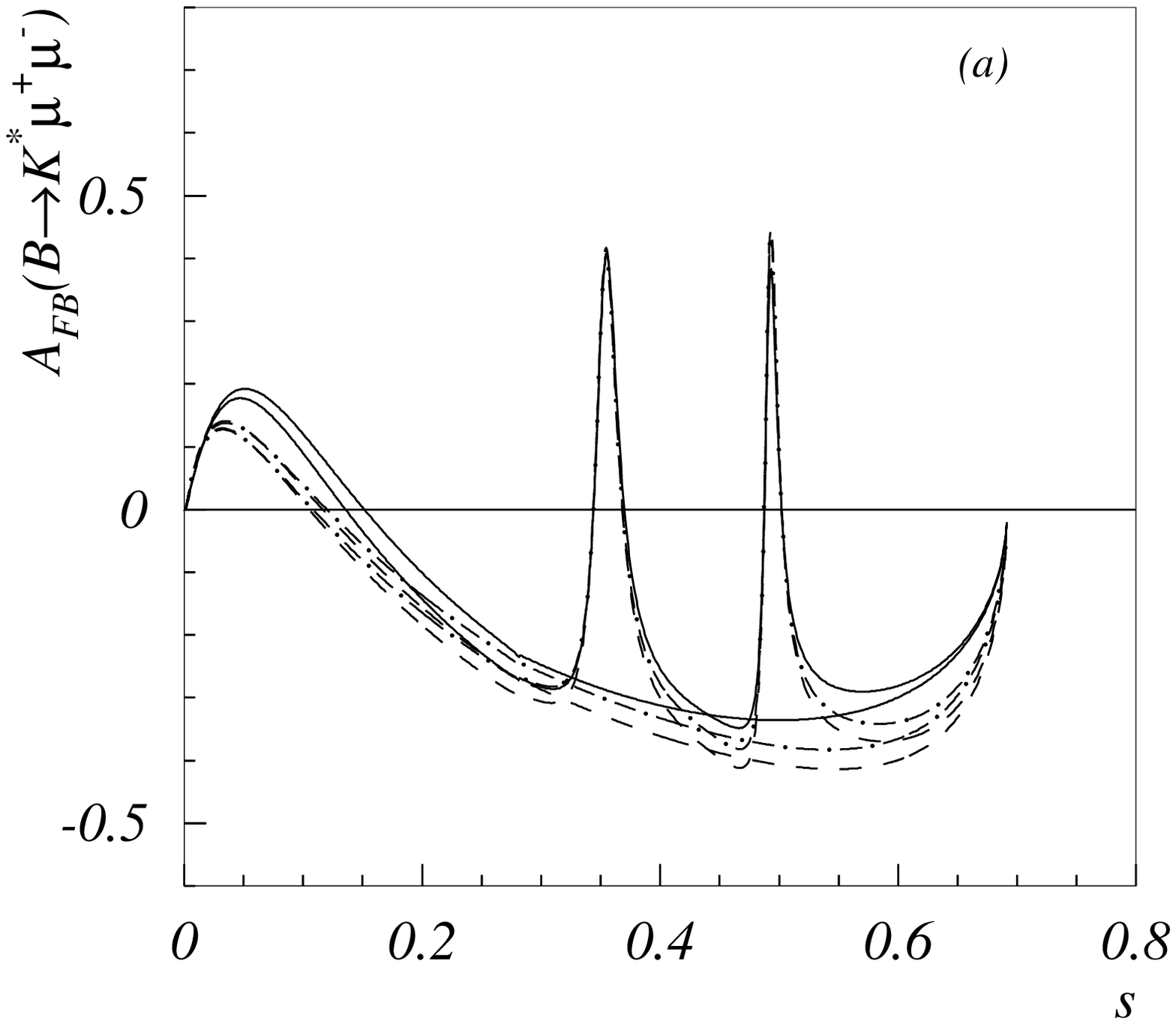}
\vskip 5.5cm
\end{figure}

\vskip 2.cm
\begin{figure}[h]
\includegraphics{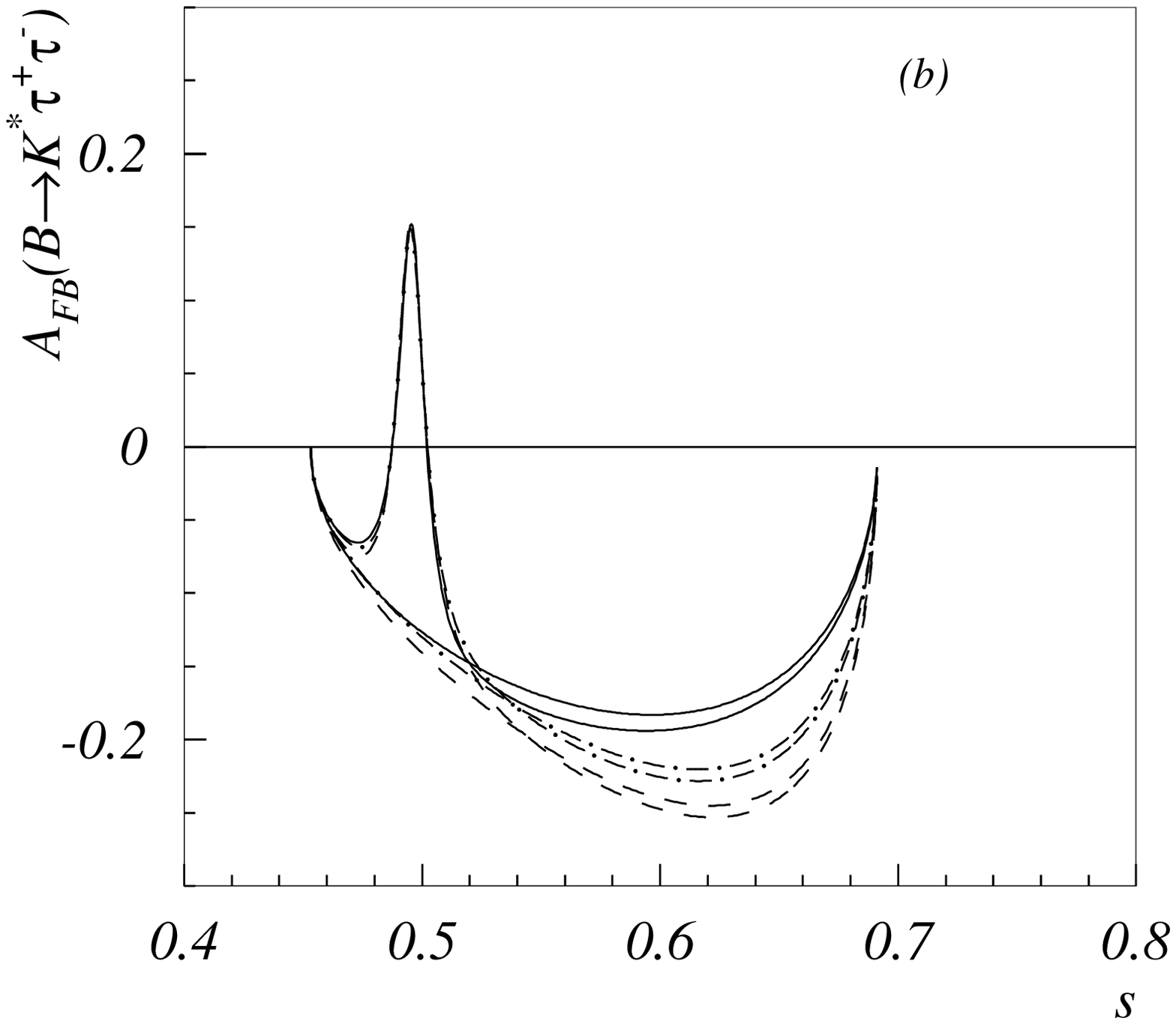}
\vskip 8.cm
\caption{ Forward-backward asymmetries for (a) $B\rightarrow K^{*}\mu
^{+}\mu ^{-}$ and (b) $B\rightarrow K^{*}\tau ^{+}\tau ^{-}$. Legend is the
same as Figure 3. }
\end{figure}

\newpage
\begin{figure}[h]
\includegraphics{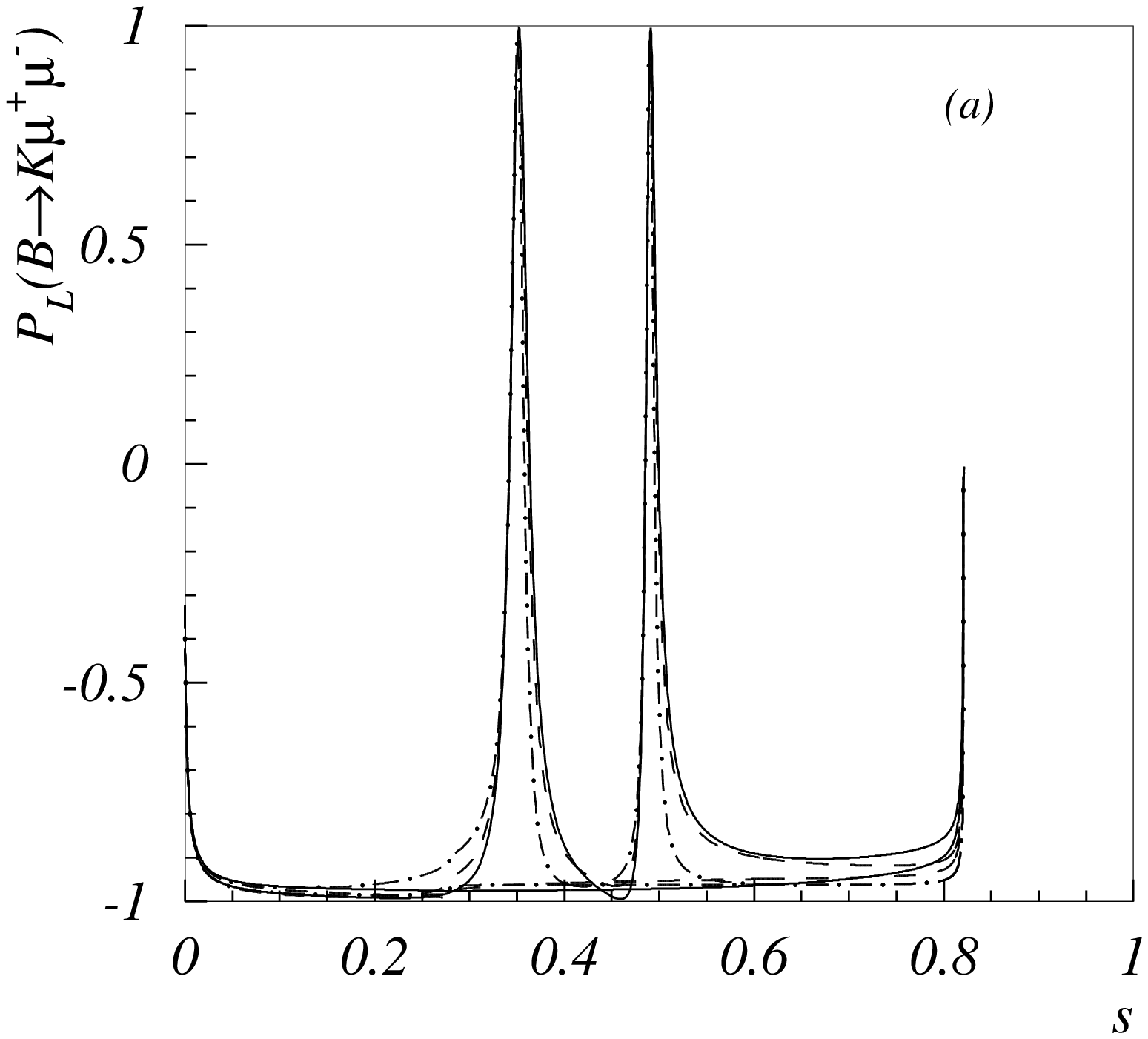}
\vskip 5.5cm
\end{figure}

\vskip 2.cm
\begin{figure}[h]
\includegraphics{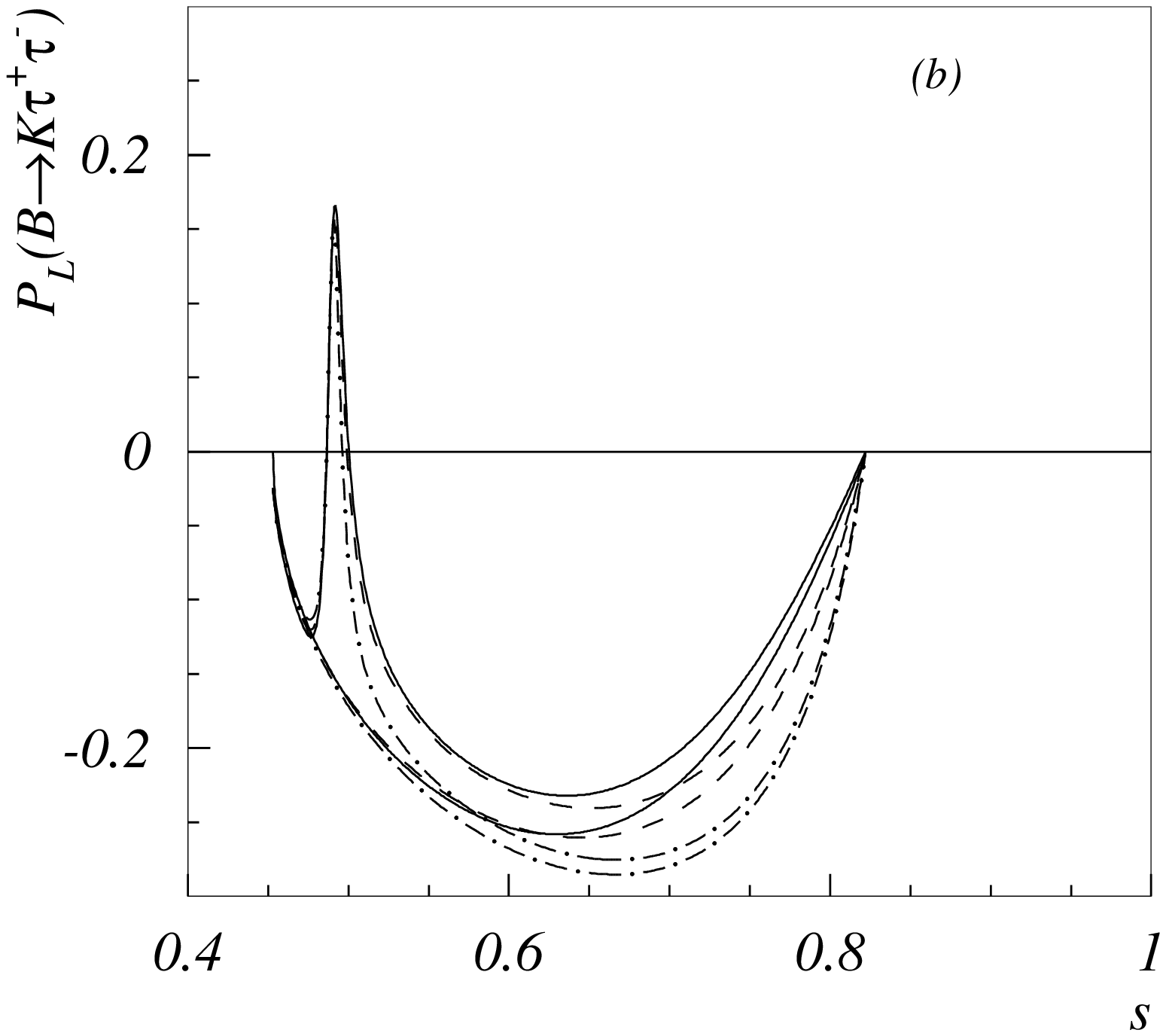}
\vskip 8.cm
\caption{ Longitudinal polarization asymmetries for (a) $B\rightarrow K\mu
^{+}\mu ^{-}$ and (b) $B\rightarrow K\tau ^{+}\tau ^{-}$. Legend is the same
as Figure 3. }
\end{figure}

\newpage
\begin{figure}[h]
\includegraphics{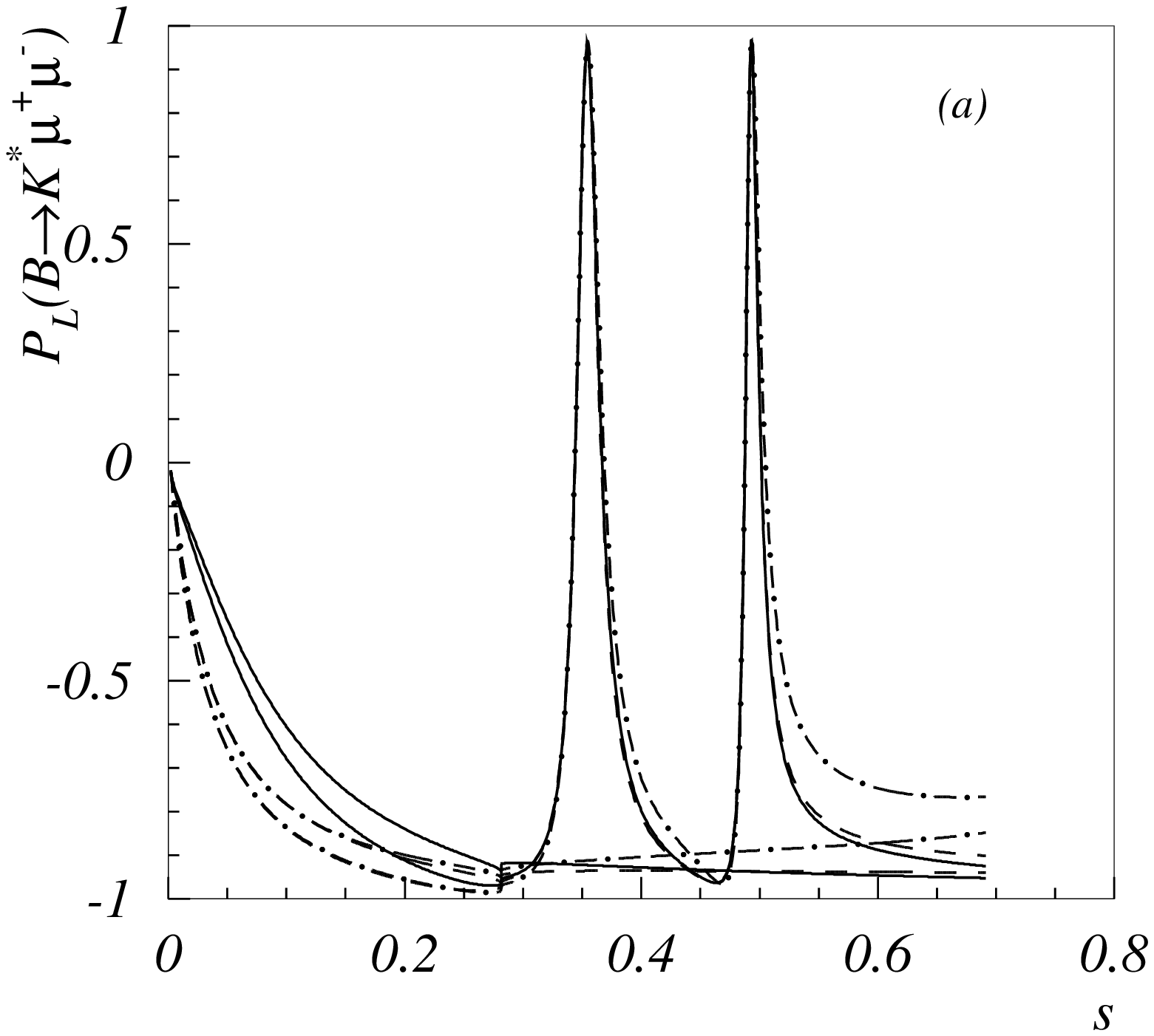}
\vskip 5.5cm
\end{figure}

\vskip 2.cm
\begin{figure}[h]
\includegraphics{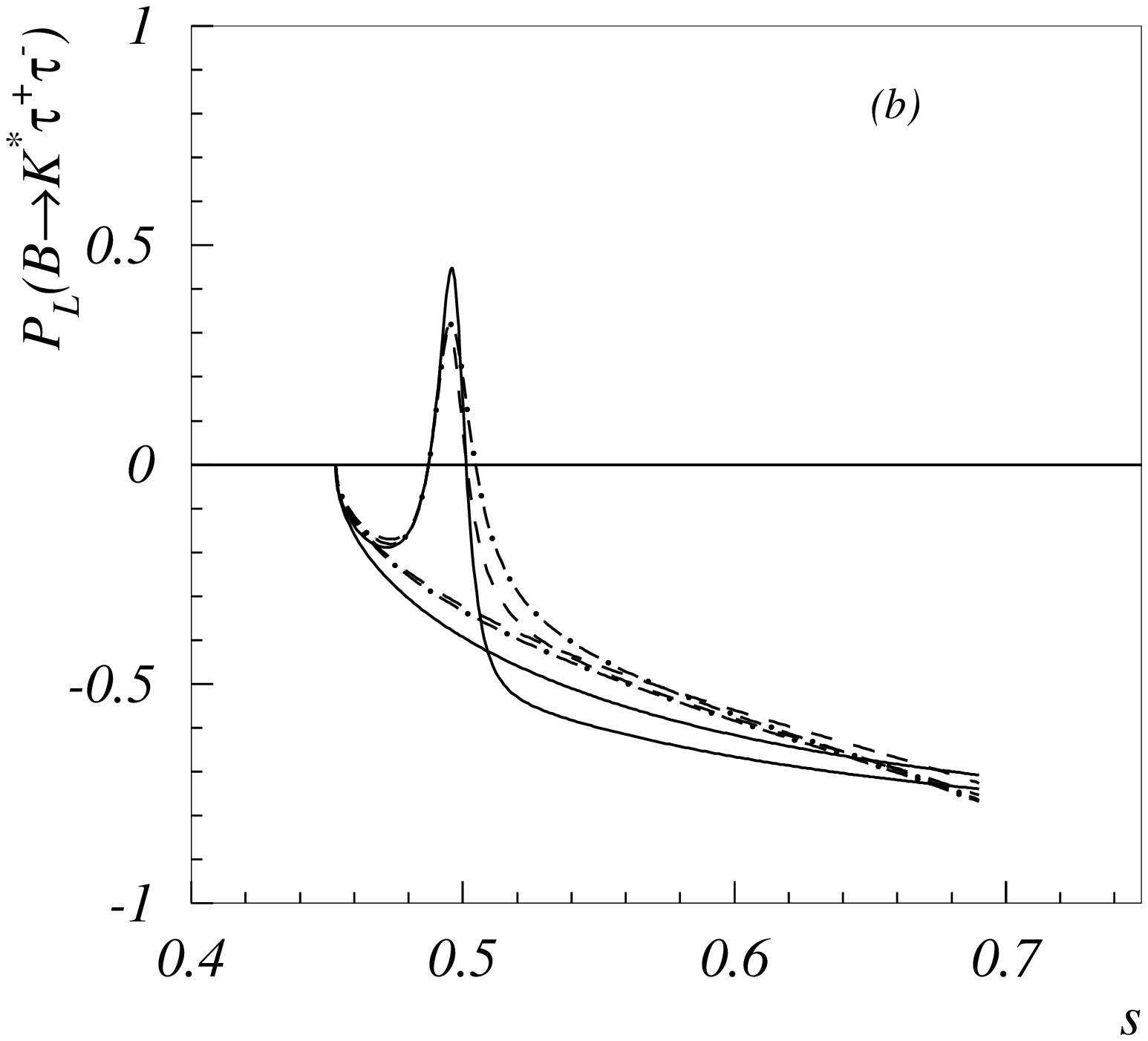}
\vskip 8.cm
\caption{ Longitudinal polarization asymmetries (a) $B\rightarrow K^{*}\mu
^{+}\mu ^{-}$ and (b) $B\rightarrow K^{*}\tau ^{+}\tau ^{-}$. Legend is the
same as Figure 3. }
\end{figure}

\newpage
\begin{figure}[h]
\includegraphics{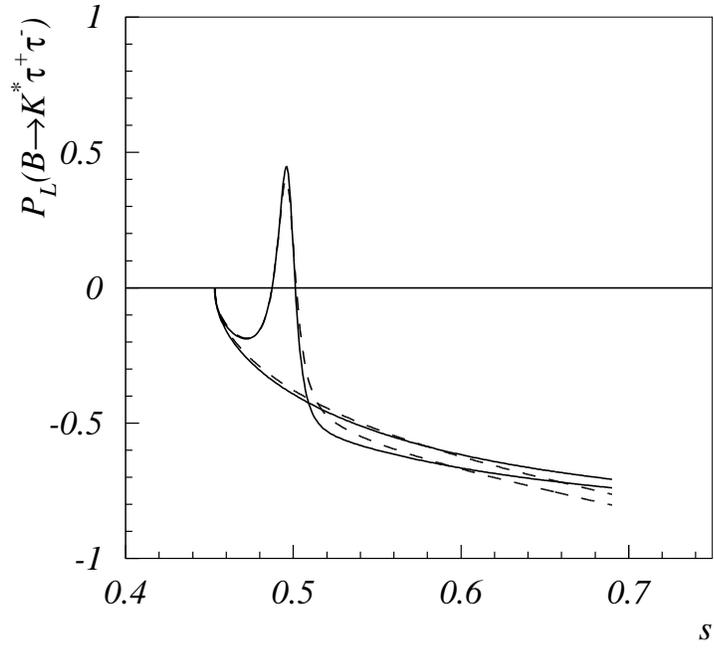} %
\vskip 13cm
\caption{ The longitudinal polarization asymmetry for $B\rightarrow
K^{*}\tau ^{+}\tau ^{-}$ in the PQCD (solid curves) and LF (dashed curves),
respectively. }
\end{figure}

\end{document}